\documentclass[iop]{emulateapj}

\usepackage{amsmath, amsthm, amssymb,amsfonts}
\usepackage{url}

\usepackage{tikz}
\usetikzlibrary{shapes, arrows, shadows}
\usetikzlibrary{decorations.pathreplacing}

\newcommand{\pic}[2][1]{\includegraphics[width=#1\linewidth]{#2}}
\providecommand{\nolinkurl}[1]{\url{#1}}

\newcommand{\oldstuff}[1]{}
\newcommand{\newstuff}[1]{#1}

\begin{document}
	
	\title{Optimal and efficient streak detection in astronomical images}

	\author{Guy Nir\altaffilmark{1}}
	\author{Barak Zackay\altaffilmark{2}}
	\author{Eran O.~Ofek\altaffilmark{1}}
	
	\altaffiltext{1}{Benoziyo Center for Astrophysics, Weizmann Institute
		of Science, 76100 Rehovot, Israel}
	\altaffiltext{2}{Institute for Advanced Study, 1 Einstein Drive, Princeton, NJ 08540, USA}

	\begin{abstract}
		Identification of linear features (streaks) in astronomical images is important for
		several reasons, including:
		detecting fast-moving near-Earth asteroids;
		detecting or flagging faint satellites streaks;
		and flagging or removing diffraction spikes, pixel bleeding, line-like cosmic rays and bad-pixel features.
		Here we discuss an efficient and optimal algorithm for the detection of such streaks.
		The optimal method to detect streaks in astronomical images is by cross-correlating the image with a template
		of a line broadened by the point spread function of the system. 
		To do so efficiently, the cross-correlation 
		of the streak position and angle is performed using the Radon transform, 
		which is the integral of pixel values along all possible lines through an image. 
		A fast version of the Radon transform exists, which we here extend to 
		efficiently detect arbitrarily short lines.
		While the brute force Radon transform requires $\mathcal{O}(N^3)$ operations for a $N\times N$ image, 
		the fast Radon transform has a complexity of $\mathcal{O}(N^2\log(N))$. 
		We apply this method to simulated images, recovering the theoretical signal-to-noise ratio, 
		and to real images, finding long streaks of low-Earth-orbit satellites
		and shorter streaks of Global Positioning System satellites.
		We detect streaks that are barely visible to the eye, out of hundreds of images, 
		without a-priori knowledge of the streaks' positions or angles. 
		We provide implementation of this algorithm in Python and MATLAB. 
		
	\end{abstract}
	
	
	\section{Introduction}
		
	Optimal and efficient detection of straight lines in astronomical images is critical for
	addressing two complimentary problems:
	detecting objects moving fast relative to the exposure time and 
	flagging or removing unwanted moving objects and linear artefacts 
	to separate them from other transients. 
	Modern optical surveys cover an increasingly large fraction of the sky in each image 
	at increasingly greater depths 
	(e.g., the Zwicky Transient Facility (ZTF; \citealt{Zwicky_transient_facility_Bellm_Kulkarni_2015}); 
	the Large Synoptic Survey Telescope (LSST; \citealt{large_synoptic_survey_telescope_Ivezic_2007})). 
	Consequently, a large fraction of the images from such surveys will be affected by cosmic rays, 
	satellites, space debris, aircraft, asteroids and more. 
	Additionally, image artefacts, such as charge bleeding or diffraction spikes, appear in many images. 
	These objects and artefacts will generally appear as linear features,
	contaminating the analysis.
	Even low intensity streaks with a per-pixel brightness that is comparable to the noise level
	could generate many false-positives when searching for point-source transients.
	For example, a streak with a signal-to-noise ratio ($S/N$) per resolution element of $2\sigma$ (twice the noise RMS)
	can promote any incidental $3\sigma$ fluctuations in its path to the level of a $5\sigma$ false detection.
	In fact, a $2\sigma$ level streak in a $1k\times 1k$ image 
	will generate a few times more $5\sigma$ false alarms
	than random fluctuations of the background. 
	Streak detection may be required when searching for
	space debris, faint asteroids or near-Earth objects~\citep{near_earth_objects_Graves_2016, streak_detection_PTF_Waszczak_2017}.
	
	Several solutions to the problem of streak detection have been suggested, in four general classes:
	simple source detection, computer vision code, machine learning algorithms, and template fitting to line shapes. 

	Employing simple source detection (e.g.,~using Sextractor; \cite{sextractor_Bertin_1996}) 
	and restricting it to sources with elongated shapes was implemented 
	by, e.g.,~\cite{streak_detection_PTF_Waszczak_2017}. 
	This method is not sensitive to
	streaks with a $S/N$ per resolution element less than 
	about 5 times the image background noise, 
	without dramatically increasing the false-alarm rate. 
	
	The use of computer vision techniques 
	is attractive because many tools are available from problems solved for natural images, 
	e.g.,~using \oldstuff{edge detection and} voting procedures such as the Hough transform \newstuff{on binary images, after applying edge detection, followed by 
    thresholding of individual pixels}~\citep{streak_detection_Hough_transform_Duda_1972, streak_detection_Hough_transform_Cheselka_1999}.
	This approach, which has been popular in recent years 
	(e.g.,~\cite{streak_detection_cosmic_rays_Keys_2010, streak_detection_pipeline_Virtanen_2014, streak_detection_ASTRIDE_Kim_2016, streak_detection_rectangles_Bektesevic_2017}),
	\newstuff{has been successful in finding streaks in crowded fields and in images 
		with diffuse light sources. 
		This is not surprising, as these methods are based on detection of linear features in natural images. 
		However, when searching for very faint streaks in uncorrelated background, 
		e.g., sparse fields or well subtracted images, 
		these methods have not been proven to be optimal. }
		
	\oldstuff{does not address, however, some aspects unique to astronomical images.
	For example, the detection of streaks just above the noise floor.
	Streaks that can barely be seen by the eye are often not detected by these methods, 
	as there is not enough signal in each individual pixel to pass the initial thresholding. }

	Machine learning algorithms	are useful in detecting assorted, 
	and often unknown, templates
	or for separating populations with a large number of variables. 
	In the case of streaks, the statistical model is well known;
	any machine learning code, if applied correctly, 
	can only reach the sensitivity of the known, optimal solution. 
	This method would be less sensitive if the input training set is not properly chosen. 
	We are not aware of any machine learning used in the detection of streaks.
	However, it has been used with some success to classify different types of streaks (e.g.,~diffraction spikes vs.~satellites); \cite{streak_detection_PTF_Waszczak_2017}.
	
	Since a streak has a well-defined statistical model, 
	one can define the optimal statistic for streak detection:
	matching the shape of the streak and 
	calculating a weighted sum of the pixels containing the line 
	(the matched-filter approach; \citealt{matched_filter_Turin_1960}). 
	\cite{streak_detection_maximum_likelihood_Dawson_2016} use the maximum likelihood method
	to calculate the optimal
	statistic for the detection of streaks. 
	For uncorrelated noise this statistic is a matched-filter 
	that has the shape of a line widened by the image
	Point Spread Function (PSF). 
	For completeness, in Appendix~\ref{sec: likelihood ratio} we derive
	this result and prove it is optimal 
	using the lemma of Nymann \& Pearson.
	\newstuff{This approach can only be considered optimal under the assumptions:
	\begin{enumerate}
		\item[(a)] The streak is a straight line, and has a constant brightness;
		\item[(b)] The streak has a constant, known PSF;
		\item[(c)] The noise is uncorrelated and there are no large scale structures in the image. 
	\end{enumerate}
	Under these assumptions the filtering we present in this work is optimal\footnote{Optimal in the sense that for a given false negative rate, 
		it has the lowest false positive rate (or vice-versa).}.
		Also note that the information loss due to (a) and (b) is quadratic and hence relatively weak (see, e.g., \citealt{Coaddition1_Zackay_2017}).
	}
	
	One way to perform the filtering is to
	cross-correlate the image with its PSF
	and then calculate integrals along all the lines crossing the image. 
	This integration process can be expressed as a Radon transform, 
	where each point in Radon-space 
	represents the sum of pixels from the original image
	along a specific slope and intercept (\citealt{Radon_transform_1917}; translated: \citealt{Radon_transform_1986}). 
	\newstuff{When both are applied to gray-scale digital images,
		the Radon transform and the Hough transform are mathematically equivalent~\citep{Hough_transform_review_Illingworth_1988}. 
		The distinction is made primarily in the preprocessing of the images. 
		Originally, the Hough transform was performed on binary images, 
		that are themselves the product of edge detection algorithms. 
		The Radon transform, on the other hand, was originally defined 
		on continuous, gray scale images. 
		We adopt the Radon transform in this work to emphasize 
		that the raw, gray-scale pixel values are added. 
		In both approaches, thresholding is used to find real streaks and 
		reject false detections. 
		In the edge detection and Hough transform method 
		the threshold is compared to each pixel value, 
		while in the Radon method the threshold is 
		compared to the sum of pixel values. }

	Calculating the Radon transform directly is computationally intensive. 
	A blind search on an $N\times N$ sized image 
	needs to scan all streak starting positions and all angles 
	while each integration requires summing $\sim N$ pixels.
	Such calculations require $\mathcal{O}(N^3)$ additions, 
	and even more calculations are needed when matching templates of different lengths and start/end points. 
	One way to approach this problem is to make multiple templates 
	for different line lengths and orientations, 
	and then to cross-correlate the image with all of them \citep{streak_detection_match_filters_Schildknecht_2015}. 
	Another approach is to perform the Radon transform using the Fast Fourier Transform (FFT) in polar coordinates, 
	and to use GPUs to speed up the interpolation step between polar and cartesian coordinates 
	\citep{streak_detection_GPUs_Zimmer_2013, streak_detection_Radon_convolution_GPU_Andersson_2015}.

	Finally, a different approach to overcoming the computational burden, 
	which we describe here, 
	is the Fast Radon Transform (FRT). 
	In astronomy, this method was developed (under different names) to solve equivalent problems:
	folding periodic signals~\citep{fast_folding_analysis_Staelin_1969};
	and de-dispersing pulsar signals in radio astronomy~\citep{dedispersion_tree_Taylor_1974}.
	\cite{fast_dispersion_measure_Zackay_2014} expanded the latter method to non-linear features (e.g.,~dispersion measure). 
	A fast approach to the Radon transform and its inverse, as well as expansions to using 
	other operators besides summation, is presented in~\cite{fast_radon_transform_inverse_generalizations_Press_2006}. 
	
	In this work, we explore the uses of the FRT algorithm for detecting streaks in astronomical images. 
	We show the algorithm is a factor $N/\log(N)$ faster than brute force.
	We further extend this method
	to scan for short streaks without a dramatic increase in runtime, 
	and use it to detect multiple streaks in the same image. 
	Finally, we provide code in Python and MATLAB. 
    
    In \S\ref{sec: optimal detection} we discuss the optimal method for streak detection and calculate the maximum $S/N$ attainable 
    for the detection of streaks in astronomical images.
    In \S\ref{sec: algorithm} we describe the FRT algorithm. 
    In \S\ref{sec: pipeline} we describe step-by-step the pre- and post-processing used in conjunction with the FRT algorithm. 
    In \S\ref{sec: simulations} we test the detection algorithm on simulated images
    and demonstrate the efficiency for recovering the signal. 
    In \S\ref{sec: real data} we test this algorithm on real images. 
    In \S\ref{sec: false alarm} we estimate the false detection rate, 
    while in \S\ref{sec: code} we discuss the streak detection code repository, 
    and in \S\ref{sec: summary} we review our results and discuss possible applications. 
	
	
	
	\section{Optimal method for streak detection}\label{sec: optimal detection}
		
		Streaks in astronomical images can be modeled 
		by a straight line convolved with the PSF of the image, 
		\begin{equation}\label{eq: streak model}
			s = \xi\ell(x_1,y_1,x_2,y_2) \otimes P.
		\end{equation}
		Here, $\xi$ is the intensity of the streak in counts per unit length, 
		$\ell$ is the single-pixel width line going from coordinates $(x_1,y_1)$ to $(x_2,y_2)$, 
		$P$ is the PSF of the system and $\otimes$ represents the convolution operator in $x$ and $y$. 
		A background subtracted image with only a streak $s$ in it would be represented by:
		\begin{equation}\label{eq: streak image model}
			M = s + \varepsilon,
		\end{equation}
		where  $\varepsilon$ is the background noise, which is assumed to be Gaussian, independent and identically distributed (i.i.d). 
		We also assume the images are background-noise dominated.
		As shown in Appendix~\ref{sec: likelihood ratio},
		the optimal way to detect such streaks is to calculate the matched-filter,
		i.e.,~cross correlate image $M$ with the streak model $s$ for every possible streak 
		and compare it to some threshold $\eta$:
		\begin{equation}\label{eq: matched filter}
			\frac{M\otimes \overleftarrow{s}}{\sqrt{s\otimes \overleftarrow{s}}} > \eta,
		\end{equation}
		where $\overleftarrow{\square}$ represents coordinate reversal in $x$ and $y$, 
		turning the convolution operator into a cross-correlation operator. 
		The $s\otimes \overleftarrow{s}$ is a normalization term, 
		which allows the same threshold to be used for any streak angle and starting position. 
		For practical purposes it is easier to first cross-correlate (filter) the 
		image with the PSF and then integrate the pixel values along 
		all possible lines going through the filtered image. 
		In \S\ref{sec: algorithm} we present an efficient method for performing this calculation using the Fast Radon Transform.
		
		When integrating using the correct streak parameters,  
		the $S/N$ for detection is the square root of the sum of the squares of the $S/N$ in each pixel\footnote{
				Since $(S/N)^2$ is an additive quantity; see, e.g.,~\cite{Coaddition1_Zackay_2017}.}. 
		The $S/N$ for a streak of intensity per unit length $\xi$ and length $L$ moving along the $y$ axis,
		widened by a Gaussian PSF with a width parameter $\sigma_p$ along the $x$ axis, is
		\begin{align}
		(S/N)_\text{total}^2 &= \int_0^L \int_{-\infty}^{\infty}(S/N)_\text{pix}^2 dxdy \notag \\ 
		&= \int_0^L \int_{-\infty}^{\infty}\left(\frac{\xi}{\sqrt{2\pi\sigma_p^2}} \exp(-\frac{x^2}{2\sigma_p^2})/\sqrt{B}\right)^2 dxdy \notag \\
		&= \frac{\xi^2}{B}\frac{1}{2\pi\sigma_p^2}L\int_{-\infty}^{\infty}\exp(-x^2/\sigma_p^2)dx \notag \\
		&=\frac{\xi^2}{B}\frac{L}{2\sqrt{\pi}\sigma_p}.
		\end{align}
		The image variance per pixel is given by $B$ (e.g.,~the combined background and read noise variance). 
		The total signal-to-noise ratio is
		\begin{equation}\label{eq: total snr}
			(S/N)_\text{total} = \xi \sqrt{ \frac{L}{2\sqrt{\pi}\sigma_p B }}.
		\end{equation}
%
%
%
		From this we can find the intensity per unit length of a streak that is detected at some $S/N$:
		\begin{equation}\label{eq: signal to noise calculation}
			\xi = S/N \sqrt{\frac{2\sqrt{\pi}\sigma_p B}{L}}.
		\end{equation}
		This equation can be used to perform streak photometry. 
		The way the streak $S/N$ changes with exposure time $T$ is given by
		\begin{equation} 
			S/N \propto \begin{cases} \sqrt{T} & \text{read noise dominated} \\ \text{const} & \text {background dominated}. \end{cases}
		\end{equation}
		We define the signal-to-noise ratio per resolution element 
		to be the $S/N$ of a section of line of a length equal to 
		the FWHM of the PSF (i.e., $2.355\sigma_p$): 
		\begin{equation}\label{eq: snr fwhm}
			(S/N)_\text{FWHM}\approx 0.81\xi/\sqrt{B},
		\end{equation}
		which is approximately the $S/N$ attainable by 
		a point source search using only a PSF-matched filter.
		
		\newstuff{If the PSF of the image is not well measured, 
		or if the PSF changes during the exposure, 
		the algorithm is no longer considered optimal  
		(e.g., meteor brightness and width are not constant along the streak path, 
		see \citealt{streak_detection_comets_Bektesevic_2018}).
		However, \cite{Coaddition1_Zackay_2017} showed that the information loss 
		due to PSF errors is quadratic in the error of the PSF width.}
			
		
	\pagebreak
		
	\section{Efficient method for streak detection}\label{sec: algorithm}
		
	In this section, we present a method for calculating the Fast Radon Transform
	using dynamic programming. 
	The Radon transform is defined as the integral along all the lines 
	(of all the angles and initial positions) that go through an image: 
	\begin{equation}
		R(x_0,\Delta x) \equiv \sum_{y=0}^{N_y-1} I(x_0+\Delta x\frac{y}{N_y},y).
	\end{equation}
	Note that we chose to define the angle of the lines going through the image as the ratio 
	between $\Delta x$, the distance between the beginning and end of the streak in the $x$ direction, 
	and $N_y$, the total number of pixels in the $y$ direction. 
	This highlights the discrete nature of lines in digital images, 
	where lines starting from the same point $x_0$ differ in angle
	in uniform steps of single pixels in $\Delta x$ when crossing the entire length of the image\footnote{
		Note that for implementations that use column major order (such as FORTRAN or MATLAB)  
		the streaks should instead be integrated along the $x$ axis (using $y_0$ and $\Delta y$).
		Also, in most column major programs indices will run from 1 to $N$, not 0 to $N-1$. 
		We adopt the row major convention (e.g.,~C and Python) in this work.}.
	We integrate over all angles spanning $-45^\circ\leq \theta \leq 45^\circ$, 
	then transpose the image and calculate the Radon transform again to sum along the remaining angles. 
	
	Each point in the Radon image is the output of an ideal filter 
	for a line of single-pixel width that crosses the entire image, 
	with a specific initial position $x_0$ and angle $\Delta x$. 
	For an $N_x\times N_y$ image, the total possible initial positions is just $N_x$, 
	while the total number of distinct angles spans $-N_y<\Delta x<N_y$ in the allowed $-45^\circ<\theta<45^\circ$, 
	for a total of $2N_y-1$ angles. 
	If each streak is assumed to cross the entire image, 
	$\mathcal{O}(N_y)$ additions are required to calculate the sum for each angle and each starting point, 
	which brings the total number of calculations in the brute force Radon transform to $\mathcal{O}(N_xN_y^2)$. 
	In the following subsection, we describe an approach 
	that reduces the number of calculations to $2N_x N_y \log_2(N_y)$. 
	For typical astronomical images, this is 2 or 3 orders of magnitude faster than the brute force algorithm.
	
	\subsection{The Fast Radon Transform algorithm}\label{subsec: frt}
	
	The Fast Radon Transform is a dynamic programming algorithm, 
	that stores results from one step to be used in the next, 
	thus avoiding redundant calculations. 
	In the case of streak detection, 
	the key idea is that multiple lines, of adjacent angles, 
	will share many of the same pixels along the path of integration;
	e.g.,~lines with the same initial $x_0$ that differ by $\Delta x=1$ in an image of size $N_y$
	pass through the exact same pixels in the first half of the data $y<N_y/2$.  
	Summing those pixels only once will save many unnecessary additions. 
	
	From a top-down perspective, 
	the algorithm uses the fact that given two halves of the image, 
	for which all possible lines have already been calculated, 
	only a few more calculations are needed to combine all lines on the full image. 
	Every time two parts of the data are combined, each part already contains 
	the sums over all possible lines, but at lower resolution. 
	Because of this partitioning scheme the input image must have a size $N_y=2^m$ for integer $m$, 
	and there would be $m=\log_2(N_y)$ steps of increasing resolution. 

	From a bottom-up perspective, 
	the algorithm begins by taking pairs of rows
	and adding them either without a shift ($\Delta x=0$), 
	or with a positive or negative shift of a single pixel ($\Delta x=\pm 1$).
	We refer to this operation as \emph{shift-and-add}, see Figure~\ref{fig: algorithm 1}.
	For two rows, this covers all summations needed for all the lines in the range $-45^\circ\leq\theta\leq+45^\circ$. 
	Once the rows are shifted and summed, 
	the results are stored in a new matrix that has an additional dimension
	to store the different results for the various shifts. 
	Instead of an $N_x\times N_y$ image, 
	the result from the first step is 3 matrices of size $N_x\times N_y/2$, 
	corresponding to the shifts $\Delta x=-1,0,+1$. 
	For convenience, this is stored in a single, three-dimensional matrix of size $N_x\times N_y/2 \times 3$. 
	The first step is shown graphically in Figure~\ref{fig: algorithm 2}. 
	
	\def\rectrow #1#2#3#4#5#6{\draw [fill=white, #6] (0,0) rectangle node{#1} +(1,1) (1,0) rectangle node{#2} +(1,1) (2,0) rectangle node{#3} +(1,1) (3,0) rectangle node{#4} +(1,1) (4,0) rectangle node{#5} +(1,1)}
	
	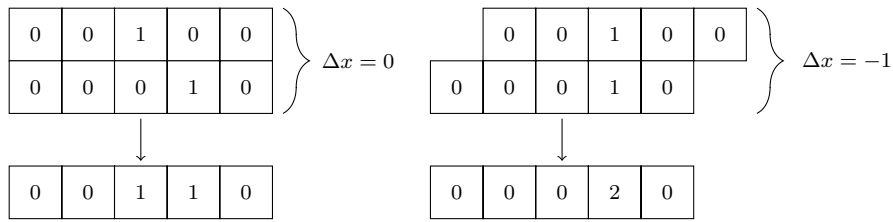
\begin{figure*}
		
		\centering
		\begin{tikzpicture}[scale=0.7]
		\rectrow 00100{yshift=3cm};
		\rectrow 00010{yshift=2cm};
		\draw [decorate,decoration={brace,amplitude=10pt, mirror}, xshift=0.2cm] (5,2) -- node[xshift=1cm]{$\Delta x=0$}(5, 4);
		\draw[->] (2.5, 1.9)--(2.5, 1.1);
		\rectrow 00110{yshift=0cm, xshift=0};
		\rectrow 00100{yshift=3cm, xshift=9cm};
		\rectrow 00010{yshift=2cm, xshift=8cm};
		\draw [decorate,decoration={brace,amplitude=10pt, mirror}, xshift=0.2cm] (14,2) -- node[xshift=1.2cm]{$\Delta x=-1$}(14, 4);
		\draw[->] (10.5, 1.9)--(10.5, 1.1);
		\rectrow 00020{yshift=0, xshift=8cm};
		\end{tikzpicture}
		\caption{The core calculation of the FRT algorithm is shift-and-add. 
			Pixel values are shown in the squares, 
			and two examples of a shift are shown for two rows of the data. 
			The signal is represented by values of 1, 
			while background noise is set to 0 for clarity. 
			In this example, the correct shift would be $\Delta x=-1$, 
			leading to an increased signal. 
		}
		\label{fig: algorithm 1}
		
	\end{figure*}
	
	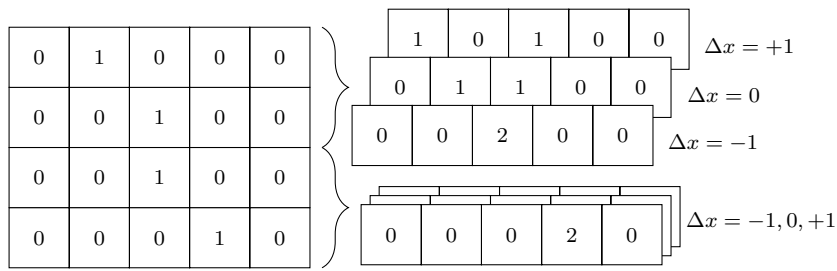
\begin{figure*}
		
		\centering
		\begin{tikzpicture}[scale=0.8]
		\rectrow 00010{yshift=0cm};		
		\rectrow 00100{yshift=1cm};		
		\draw [decorate,decoration={brace,amplitude=10pt, mirror}, xshift=0.2cm] (5,2) -- (5, 4);
		\rectrow 00100{yshift=2cm};
		\rectrow 01000{yshift=3cm};
		\draw [decorate,decoration={brace,amplitude=10pt, mirror}, xshift=0.2cm] (5,0) -- (5, 2);
		\rectrow 10100{xshift=6.3cm, yshift=3.3cm} node[xshift=0.8cm, yshift=-0.5cm]{$\Delta x=+1$};
		\rectrow 01100{xshift=6cm, yshift=2.5cm} node[xshift=0.7cm, yshift=-0.5cm]{$\Delta x=0$};
		\rectrow 00200{xshift=5.7cm, yshift=1.7cm} node[xshift=0.8cm, yshift=-0.5cm]{$\Delta x=-1$};
		\rectrow {}{}{}{}{}{xshift=6.15cm, yshift=0.35cm, scale=1};
		\rectrow {}{}{}{}{}{xshift=6cm, yshift=0.2cm, scale=1} node[xshift=1.2cm, yshift=-0.35cm]{$\Delta x=-1,0,+1$};
		\rectrow {0}{0}{0}{2}{0}{xshift=5.85cm, yshift=0.05cm, scale=1};
		\end{tikzpicture}
		\caption{The first step in the FRT algorithm. 
			Four rows of five pixels are presented
			with 1 representing the signal, with the background noise set to 0 for simplicity.
			A streak passes through the rows with an initial position of $x_0=3$ 
			and a slope corresponding to $\Delta x=-2$.			 
			The top two rows are added together using three different shifts, $\Delta x=-1,0,+1$, 
			where the top row is shifted and then added to the one beneath it. 
			A positive shift moves the top row to the left, 
			a negative one to the right,  
			and a zero-shift is just the addition of the two rows. 
			The results are stored in a slice going into the 3rd dimension. 
			The bottom two rows go through the same process and the results are stored in another slice. 
			This step is done iteratively for all pairs of rows in the image. 
			For column-major memory environments like MATLAB, the addition is run on columns
			and is performed from left to right instead of top to bottom, 
			on values of $y_0$ and $\Delta y$ instead of $x_0$ and $\Delta x$. 
			Note that all rows (not just those with a high signal) are stored; 
			the FRT calculates all possible lines through the image. 
		}
		\label{fig: algorithm 2}
	\end{figure*}
	
\begin{figure*}
		
		\centering
		\begin{tikzpicture}[scale=0.8]
		\rectrow {}{}{}{}{}{xshift=0.15cm, yshift=2.35cm, scale=1};
		\rectrow {}{}{}{}{}{xshift=0cm, yshift=2.2cm, scale=1};
		\rectrow {0}{0}{2}{0}{0}{xshift=-0.15cm, yshift=2.05cm, scale=1};
		\rectrow {}{}{}{}{}{xshift=0.15cm, yshift=0.35cm, scale=1};
		\rectrow {}{}{}{}{}{xshift=0cm, yshift=0.2cm, scale=1};
		\rectrow {0}{0}{0}{2}{0}{xshift=-0.15cm, yshift=0.05cm, scale=1};
		\draw [decorate,decoration={brace,amplitude=10pt, mirror}, xshift=0.2cm] (5,0.25) -- (5, 3.25);
		\rectrow 11010{xshift=6.9cm, yshift=3.5cm} node[xshift=0.8cm, yshift=-0.5cm]{$\Delta x=+3$};
		\rectrow 02010{xshift=6.6cm, yshift=2.75cm} node[xshift=0.8cm, yshift=-0.5cm]{$\Delta x=+2$};
		\rectrow 11110{xshift=6.3cm, yshift=2cm} node[xshift=0.8cm, yshift=-0.5cm]{$\Delta x=+1$};
		\rectrow 01210{xshift=6cm, yshift=1.25cm} node[xshift=0.7cm, yshift=-0.5cm]{$\Delta x=0$};
		\rectrow 00220{xshift=5.7cm, yshift=0.5cm} node[xshift=0.8cm, yshift=-0.5cm]{$\Delta x=-1$};		
		\rectrow 00040{xshift=5.4cm, yshift=-0.25cm} node[xshift=0.8cm, yshift=-0.5cm]{$\Delta x=-2$};		
		\rectrow 00022{xshift=5.1cm, yshift=-1.0cm} node[xshift=0.8cm, yshift=-0.5cm]{$\Delta x=-3$};
		\end{tikzpicture}
		\caption{The second step in the FRT algorithm. 
			Slices from the previous step are added together with an additional offset of $\Delta x=-2,\ldots +2$.
			Combining the offset from the previous step, larger offsets of $\Delta x=-3,\ldots +3$ can be calculated. 
			In this example, the correct offset is $\Delta x-2$, which is recovered by adding
			the two rows that have $\Delta x=-1$ from each of the two slices, with an additional shift of -1.
			The additional shift is added to compensate for the difference between the beginning and end point of the first slice. 
			From two slices with $5\times 2\times 3$ pixels, this step produces a single slice of $5\times 1 \times 7$ pixels. 
			In images with $N_x\times N_y$ pixels, after $\log_2 N_y$ iterations, 
			the final, single slice is of size $N_x\times 1 \times (2N_y-1)$, which covers 
			all values $\Delta x=-N_y,\ldots,0,\ldots,N_y$ corresponding to a slope angle of $\theta=-45^\circ\ldots,+45^\circ$.
			This slice, with dimension permuted back to $x$ and $y$, is the output Radon image. 
			The point in the Radon image with the highest signal corresponds to the correct streak parameters. 
		}
		\label{fig: algorithm 3}
	\end{figure*}
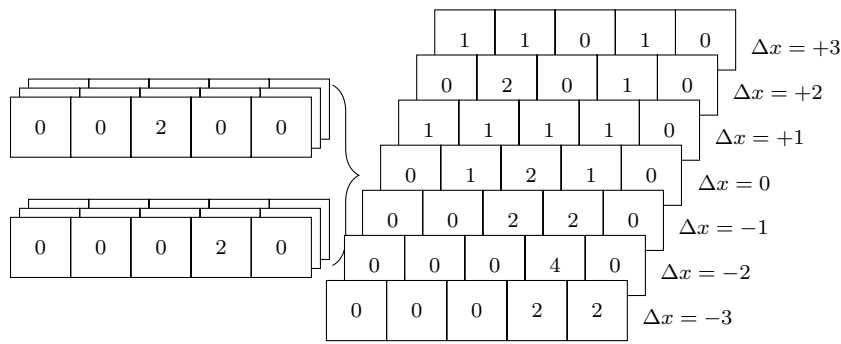

	In the second step (and all subsequent steps), 
	the results from the previous step are shift-and-added together 
	to give the integrals over larger subsections of the image. 
	When adding two pairs of rows, for example, 
	the range of possible slopes is $\Delta x=-3,\ldots, +3$, 
	which is built up by taking the right rows with the pre-existing offsets 
	and summing them, often with an additional relative shift. 
	In the second step of the algorithm, for example, 
	we can construct the integral over all $\Delta x=+2$ lines
	by shift-and-adding the two existing $\Delta x=+1$ rows from the previous step. 
	A graphical description of the results of the second step is shown in Figure~\ref{fig: algorithm 3}. 
	In each step $m$ in the algorithm, we calculate
	\begin{align}\label{eq: core frt}
		M'[i,j,k] &= M[2i, \text{fix}(j/2), k] \notag\\
	          &\qquad + M[2i+1, \text{fix}(j/2),k+\text{ceilfix}(j/2)],
	\end{align}
	where $M$ is the matrix generated in step $m-1$, $M'$ is the new matrix calculated in step $m$, 
	the index $i$ runs over $N_y/2^m$ slices of the data, 
	the index $j$ runs over all possible angles $-2^m<j<2^m$, 
	and the index $k$ runs over all $x$ values in the matrix. 
	The operators fix() and ceilfix() round fractions towards zero and away from zero, respectively. 
	Wherever $k+\text{ceilfix}(j/2)$ is outside the bounds of $M$, we add zero. 	
	
	In each subsequent step, the number of different slopes grows while the length of the $y$ axis of the data shrinks. 
	At step 2, for example, we will be left with a matrix of size $N_x\times N_y/4 \times 7$. 
	For step $m$, we will have a matrix of size $N_x\times N_y/2^{m}\times(2^{m+1}-1)$.
	Note that the \emph{passive axis} ($x$ in this case) 
	does not change size during the run,  
	as rows of equal size are always added.
	It is generally faster to run the program on contiguous blocks of memory. 
	For software that store data in column-major order (such as MATLAB), it is preferable 
	to add columns and leave the $y$ axis intact while summing and halving the data along the $x$ axis. 
	For row-major order software (like C or Python), it is more efficient to sum rows and leave the $x$ axis unchanged. 
	Another important limitation is that the input data size must be an integer power of 2
	along the \emph{active axis} (for column-major order, the $x$ axis; for row-major order, the $y$ axis). 
	If the data size is not an integer power of 2 it should be zero padded. 
	Finally, the resulting matrix is larger than the input, 
	and will be sized $N_x\times(2N_y-1)$, 
	which corresponds to all possible lines in the range $-45\leq\theta\leq+45^\circ$ 
	with a resolution of $-N_y\leq\Delta x\leq+N_y$. 
	The remaining angles can be probed by
	performing the same transformation on the transposed image.

	\section{Step by step description of the algorithm}\label{sec: pipeline}
	
	We present a summary of the steps taken for streak detection. 
	Discussion of each step is provided in the following subsections. 
	
	\begin{itemize}
		\item Subtract a reference image or remove all point sources.
		\item Subtract the background.
		\item Calculate the variance of the image.
		\item Cross correlate with the PSF of the image. 
		\item For each image or variance image, use zero padding as required. 
		\item Perform FRT on the variance map and its transpose.
		\item Multiply the Radon variance by $\sqrt{\sum_x P(x)^2}$ 
		to compensate for not filtering the variance image, 
		and by $\max(|\cos\theta|, |\sin\theta|)$ 
		to compensate for the lower correlation of the pixels in diagonal lines. 
		\item Perform FRT on the image and its transpose. 
		\item Calculate the $S/N$ image by dividing each pixel in the Radon image by the square root of the corresponding pixel in the Radon variance image.
		\item Exclude areas of the $S/N$ image corresponding to vertical or horizontal lines, 
		      to avoid sensor artefacts if needed. 
		\item Locate the global maximum in the $S/N$ image and save the coordinates and $S/N$ values if they exceed the threshold. 
	\end{itemize}

	The final three steps can be applied on partial transformations (i.e.,~in each logarithmic step of the FRT) 
	to find short streaks. 
		
	\subsection{Preparing the matrix for FRT}\label{subsec: preparing matrix}
	
	The basic implementation only calculates shifts corresponding to the range $-45\leq\theta \leq 45^\circ$.
	To calculate line integrals on the remaining angles, we perform another FRT on the transpose of the image. 
	It is possible to run the FRT on the same image, in place, 
	but the summations along columns is very ineffective 
	since memory is stored along rows (and vice versa for column-major order applications). 
	It is usually faster to make a transposed copy of the image
	and then run the FRT along rows of the transposed image. 

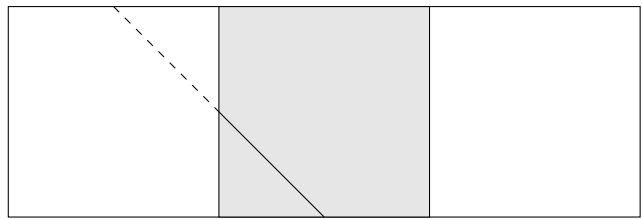
\begin{figure}
		\centering
		
		\begin{tikzpicture}[scale=1.4]
		\draw (0,0) rectangle (6,2);
		\filldraw [fill=black!10] (2,0) rectangle (4,2);
		\draw [dashed](1,2) -- (2,1);
		\draw (2,1) -- (3,0);
		\end{tikzpicture}	
		
		\caption{A cartoon of an image and zero padding. 
			The gray area is the original image, 
			with the solid line representing the measured streak. 
			Since the line begins outside the image, 
			it falls outside the Radon transform of the original image. 
			The white areas are zero padding required so that the streak, 
			and its continuation, marked by a dashed line, 
			would be within the resulting Radon image. 
			This padding of the passive axis is done in addition 
			to padding the active axis to be an integer power of 2. 
		}
		\label{fig: expand matrix}
	\end{figure}

	Input images should be zero padded to a power of 2 in the active dimension  --
	i.e.,~in the $y$ direction for row-major order.
	In addition, the passive dimension should also be zero padded by the length of the active dimension;
	e.g.,~if summing along the $y$ axis, the image should be padded to the size $(N_x+2N_y)\times N_y$.  
	This is done so that lines that cross only the corner of the image are also accounted for (see Figure~\ref{fig: expand matrix}). 
	If a line's starting point is outside the image given to the FRT, 
	the integral result over that line would be outside the Radon image. 
	Expanding the matrix beforehand increases the size of the Radon image to include all possible lines. 
	However, this expansion is not needed when using the short-streak detection method presented in \S\ref{subsec: short}. 
		
	\subsection{Preprocessing the images}\label{subsec: preprocessing}
	
	Before using the FRT algorithm to find streaks, 
	some actions should be taken to maximize the sensitivity for finding streaks.
	Any stars or bad pixels in the image should be removed,
	either by cutting small stamps around each point source 
	or by using image subtraction of a reference image, 
	ideally by a method that does not generate correlated noise (e.g.,~\citealt{Subtraction_Zackay_2016}). 
	Any residual point sources will translate to lines in the Radon image.
	Such point sources will be spread over many pixels in the Radon image, 
	but if the original sources are bright enough, they may still overpower any real streaks 
	or simply register as false positives. 
	\newstuff{Other sources of light that may interfere with streak detection include, 
		e.g., extended objects and scattered light in the telescope. 
		If the extended source is bright and does not extend over large areas of the image, 
		it can be easily identified and masked before applying the algorithm. 
		If bright sources also have diffraction spikes or pixel bleeding, 
		these artefacts would not be removed by cutting out small regions around the sources or by image subtraction.  
		They will, however, be correctly detected (and subsequently removed) by the streak finding algorithm. 
	}
		
	The image should be background subtracted. 
	Any residual bias will be coadded along lines that can be thousands of pixels long, 
	and even small biases can cause a significant false-positive result. 
	Negative bias (from over-subtracting) may result in reduced sensitivity and false-negative events. 	
	\newstuff{These requirements are not unique to the FRT algorithm, or to streak detection in general. 
			Subtraction of faint, extended light sources is necessary for, 
			e.g., transient search or precision photometry. 
			Efforts should be made to subtract this light using reference images, 
			although this is not always possible (e.g., scattered light). 
			The streak detection sensitivity can be limited by diffuse light, 
			like any other measurements made on the images. 
			If a streak has local brightness that is lower than the diffuse light pollution, 
			it would also be too faint to affect any other measurements.
			The streak detection threshold for any set of images 
			should therefore be adjusted based on each dataset, 
			as described in \S\ref{sec: false alarm}.
		}
	
	When looking for astronomical sources (e.g.,~that are above the atmosphere), 
	the image should be cross-correlated with its PSF before applying the FRT. 
	The PSF is estimated from point sources in the image, 
	either by directly measuring its shape or by 
	approximating the PSF with a 2-dimensional Gaussian with the width of the average PSF.
	Several software packages have been proposed for extracting the PSF from an image 
	(e.g.,~\citealt{extract_psf_PSFEx_Bertin_2013, extract_PSF_Mancone_2013}). 
	Exact knowledge of the PSF is not required, as errors in the PSF 
	have only a second order effect in the recovered information, $(S/N)^2$ (see \citealt{Coaddition1_Zackay_2017}).

	The FRT is equivalent to cross-correlating the image with all possible lines, 
	while filtering with the PSF is another cross-correlation.
	The order of the two operations is commutative, 
	and it is simpler to perform the PSF filter on the image rather than on the Radon image. 
	For artefacts and cosmic rays, which do not share the PSF of the system
	but often appear as lines with single pixel width, 
	the PSF filter will reduce the sensitivity of the search. 
	We propose performing the FRT on filtered images to find astronomical sources, 
	and on unfiltered images to find cosmic rays and artefacts. 
	The difference between these two procedures, for a given streak, 
	can provide an indication of whether it is an astronomical source or an artefact
	(see details in, e.g., \citealt{Subtraction_Zackay_2016}).
	\newstuff{If the streak source does not share the PSF of the image, 
		or has a variable PSF, it will be inherently harder to detect (see \S\ref{sec: optimal detection}).}

\begin{figure}
		\centering
		\pic[1]{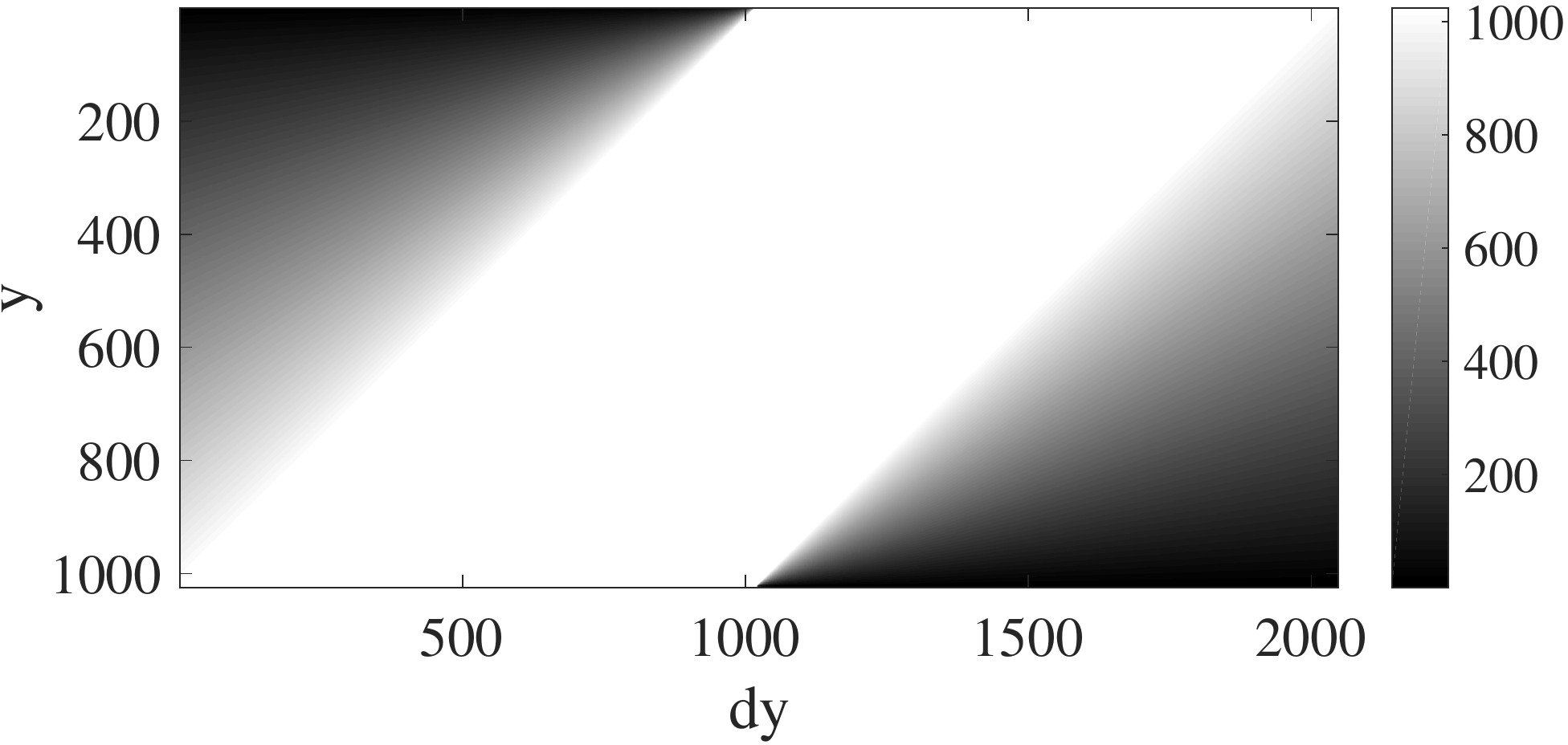}
		\caption{Radon transform of a uniform variance map for a $1024\times1024$ pixel image. 
		         Different areas of Radon space have different weight
		         corresponding to different lengths of streaks. 
	             Dividing by the square root of this image normalizes 
	             each point in the Radon image. 
	             If the overall noise variance is not unit valued, 
	             the Radon variance map can be scaled linearly by the variance value. 
	             If the image variance is not uniform, 
	             the specific variance map should be Radon-transformed 
	             and the result used instead of a uniform Radon map.
         		}
         \label{fig: radon variance}
	\end{figure}

	Finally, it is necessary to perform a Radon transform on the variance of the image, 
	and on the variance of the transposed image, so that the Radon images can later be normalized 
	to units of signal-to-noise ratio. 
	If the input image matrix is $M$, and the variance for that image is $V$, 
	we can find the significance map $S$ by using pixel by pixel division: 
	\begin{equation}\label{eq: radon snr}
		\tilde{S} = \frac{\tilde{M}}{\sqrt{\tilde{V}}},
	\end{equation}
	where the Radon transform is represented by $\tilde{\square}$. 
	Note that $V$ is either the variance of the image, filtered by the PSF, 
	or the unfiltered variance, multiplied by $\sqrt{\sum_x P(x)^2}$. 
	The variance image, which represents the variance along lines of different angles, 
	should also be multiplied by a geometric factor  $g(\theta)=\max(|\cos\theta|, |\sin\theta|)$, 
	to take into account the lower correlation between pixels in diagonal lines, as compared to vertical or horizontal lines.  
	Even if the input image has uniform variance of a known value, 
	it is necessary to perform at least one transformation on a uniform map of the same size as the images, 
	and one on the transpose of that map, 
	since different points in the Radon image will pass through a different number of pixels,
	and will have substantially different noise statistics. 
	An example of a Radon transform of a uniform variance map of unit value is shown in Figure~\ref{fig: radon variance}, 
	where the strong difference in intensity highlights the difference in the lengths of streaks across the Radon image. 
	
	\subsection{Short streaks}\label{subsec: short}
	
	It is somewhat more difficult to detect
	short streaks that cover only part of the image 
	(i.e.,~streaks that start or end inside the frame).
	An integration over full lines, as in the Radon transform, 
	will include more noisy pixels that contain none of the signal. 
	Ideally, the template should have the same start and end points as the streak. 
	In brute force calculations, integration of length $N$ can use the results from 
	a line of length $N-1$.
	In this case there is still need to test either all end points or all starting points
	for every intercept and slope. 

	To address short streaks, we modify the FRT algorithm 
	so that in each logarithmic step $m$, 
	after the shift-and-adding of all layers of a certain size, 
	the intermediate results, or \emph{partial Radon images}, 
	are scanned for a global maximum. 
	In these partial transforms,
	 the data is arranged in a 3D matrix (see \S\ref{sec: algorithm}),
	the first axis still representing $x_0$,
	the second representing $y$ in jumps of length $2^{m}$,
	and the third axis representing $\Delta x$. 
	From the position in this 3D matrix, 
	the line parameters can be reconstructed, 
	while $2^m/\cos\theta$ or $2^m/\sin\theta$ give an estimate 
	(up to a factor of two) of the length of the streak. 
	Maximum sensitivity is recovered for streaks 
	that happen to fall exactly on the boundaries of the data subsets
	(where the data is partitioned by powers of two).
	For streaks that start and end exactly in the middle of the subsets, 
	the $S/N$ will be reduced by a factor of two.
	This loss of sensitivity can be avoided 
	by triggering on a lower threshold 
	and filtering the candidates with a brute force scan 
	on streak angle, position and length parameters 
	that are close to the detection parameters. 
	Lowering the threshold will generate a few more false detections per image.
	However, filtering these candidates by scanning over a small parameter space 
	should not dramatically change the runtime. 
	
	In each step, the partial Radon transform must be divided by 
	the square root of the respective partial Radon transform 
	of the variance image (as in Equation~\ref{eq: radon snr}).
	Once the maximum $S/N$ is found for one step, 
	it can be compared to the $S/N$ found for the following steps, 
	which integrate larger subsets of the data. 
	As the algorithm proceeds to larger datasets, 
	the step where the $S/N$ peaks is the best estimate for the length of the streak. 
	To speed up calculations and avoid false detection of point source transients, 
	we suggest running this search only after reaching a predefined step $m_0$ in the algorithm, 
	and saving only streaks with a $S/N$ that improved when going to step $m_0+1$. 
	The choice of $m_0$ is done so that the minimal length of the streak is $2^{m_0+1}$. 
	
	\subsection{Multiple streaks}\label{subsec: multi}
		
	In applications where multiple streaks are expected in an image, 
	we suggest applying the same search algorithm iteratively, 
	removing the highest $S/N$ streak from the input image (replacing it with zeros\footnote{
	            For optimal detection sensitivity, the same position in the variance of the image 
                should be replaced with zeros, and a new Radon variance map produced for finding subsequent streaks. 
                This is not implemented in our code.})
	and then running the FRT again on the subtracted image. 
	This method is more robust than trying to find multiple, separate, local maxima in the Radon image, 
	and does not make the search considerably slower if we assume most images
	will not have more than one or two streaks. 
	The width of the streak in the input image (that has been filtered with the PSF)
	is around twice as wide as the original streak, 
	and we find that for most applications, removing 3 times the PSF width parameter $\sigma_p$
	on either side of the found position (in both $x_0$ and $\Delta x$ parameters), 
	is enough to prevent most streaks from being detected multiple times. 
	
	\subsection{Excluding areas in Radon space}\label{subsec: exlcusion}
	
	Some noise sources are limited to rows or columns of the detector (e.g.,~sCMOS detector line noise). 
	These lines may be common in many or all images, and generate many false alarms, 
	sometimes even overpowering real streaks in the image. 
	Since these lines appear in specific angles it is easy to discard a limited area in the Radon image
	around $\theta=0$ (or $\theta=90^\circ$ in the transposed Radon image). 
	To exclude vertical lines, we simply set to zero some rows in the Radon image around $\Delta x=0$.
	For horizontal lines, we do the same for the transposed Radon image. 
	
	
	\pagebreak
	
	\section{Simulations}\label{sec: simulations}
	
	\begin{figure*}
		
		\centering
		\pic[0.85]{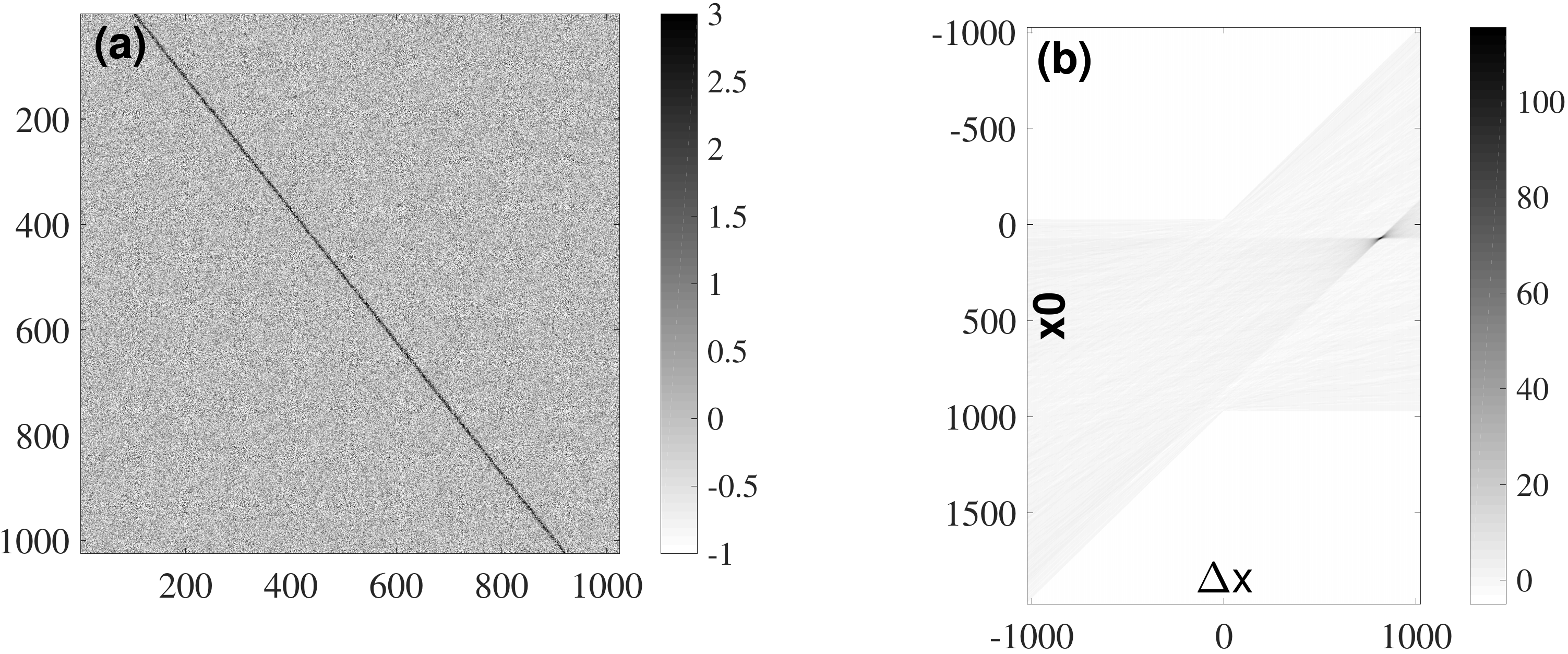}
		\caption{A bright, simulated streak and the resulting Radon image.  
			The image gray-scale is inverted such that black is bright.
			(a) The algorithm recovers the correct streak angle and position, 
			$x_0=102$ and $\theta=51.3^\circ$. 
			The simulated streak has an intensity of $\xi=10$ counts per pixel, 
			a PSF with width of $\sigma_p=2$ pixels and a noise variance of $B=1$, 
			equivalent to a $S/N\approx 8$ per resolution element. 
			This streak should be easily detected with a $S/N\cong 151.7$. 
			(b) The bright peak in the normalized Radon image gives a detection $S/N=149.2$.
		}
		\label{fig: example simulation bright}
		
	\end{figure*}

	\begin{figure*}
		
		\centering
		\pic[0.85]{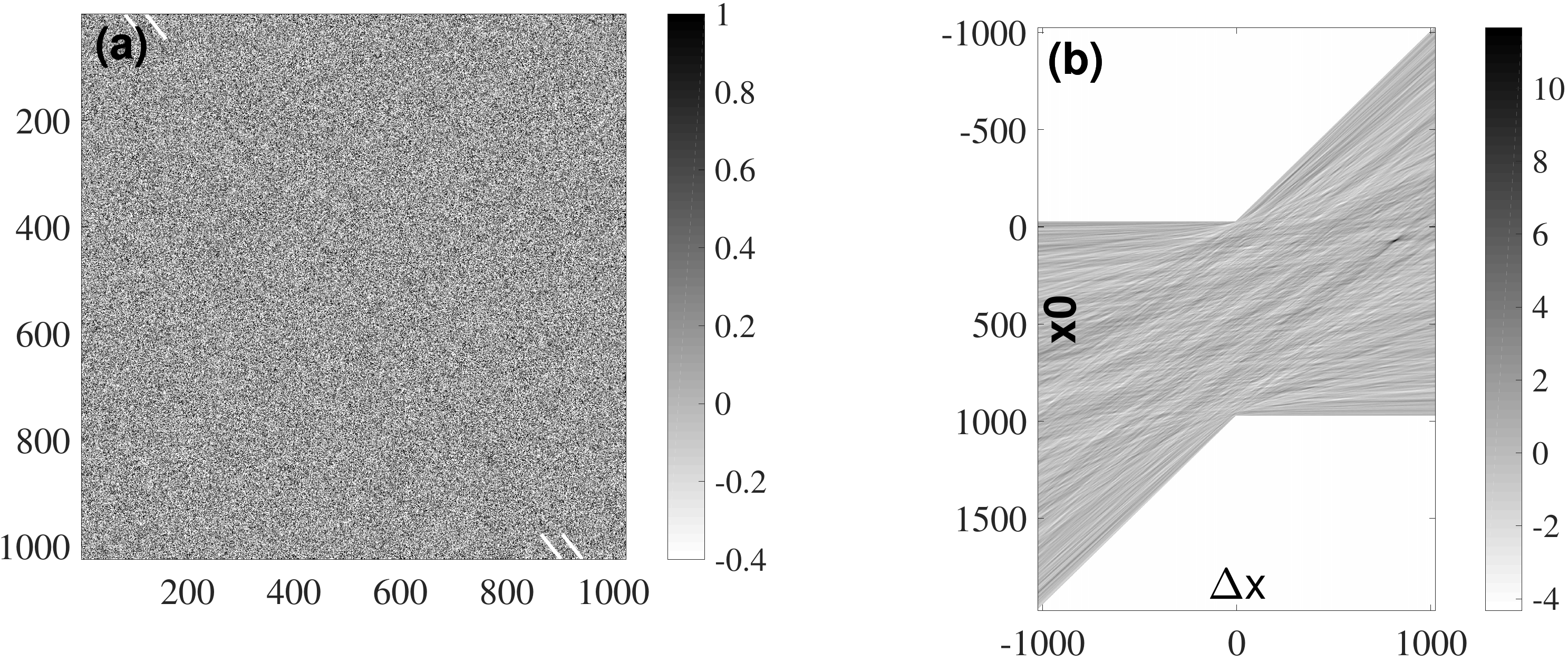}
		\caption{A faint, simulated streak and the resulting Radon image.
			(a) The dashed white lines show the beginning and end of the 
			recovered streak with an angle and position 
			$x_0=103$ and $\theta=51.4^\circ$. 
			The simulated streak has an intensity of \newstuff{$\xi=0.6$}\oldstuff{$\xi=1$} counts per pixel,
			a PSF with a width of $\sigma_p=2$ pixels and a noise variance of $B=1$, 
			equivalent to \newstuff{$S/N\approx 0.51$}\oldstuff{$S/N\approx 0.8$} per resolution element. 
			The streak is almost invisible to the eye. 
			(b) The peak in the normalized Radon image gives the correct streak coordinates. 
			The detection is made with \newstuff{$S/N\approx 8.66$}\oldstuff{$S/N\approx 15$} (while the theoretical \newstuff{$S/N\cong 8.1$}\oldstuff{$S/N=15.2$}).
		}
		\label{fig: example simulation faint}
		
	\end{figure*}

	\begin{figure*}
		
		\centering
		\pic[0.85]{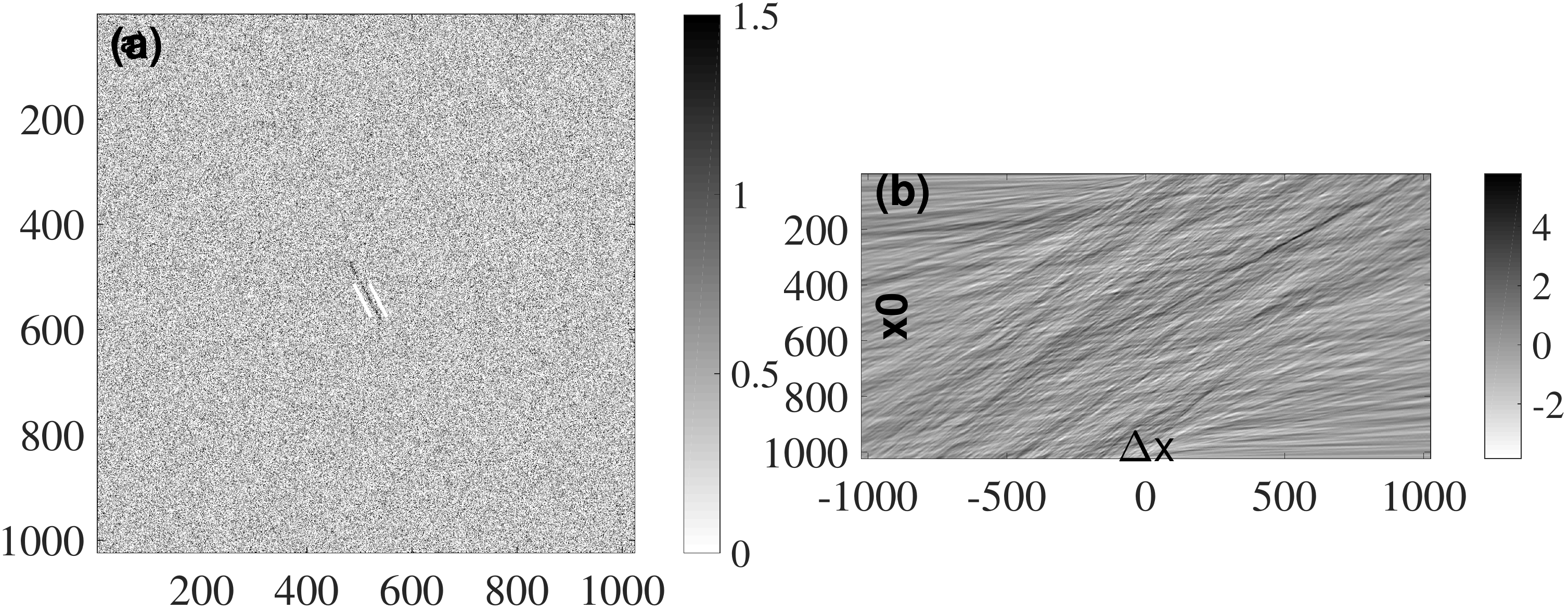}
		\caption{A short, simulated streak and the resulting Radon image.
			(a) The dashed white lines show the recovered streak position and angle,
			$x_0=231$ and $\theta=62^\circ$, and the streak length of $L=72$ pixels.
			The simulated streak has an intensity of $\xi=4$ counts per pixel, 
			a PSF with a width of $\sigma_p=2$ pixels and a noise variance of $B=1$, 
			or a $S/N\approx 3.5$ per resolution element. 
			This streak should be detectable with a $S/N=17$. 
			(b) The peak in the final, normalized Radon image gives a $S/N\approx 5.7$,
			which is lower than the detection $S/N$, since the final Radon image includes integrals over long streaks. 
			In this case, the best (short-streak) detection has a $S/N\approx 13$ for a streak with a length $L=72$ pixels. 
			This detection includes some pixels outside the streak end points, 
			and misses some of the streak pixels that are outside the integration bounds, 
			since the algorithm integrates only over rows of integer powers of 2.
		}
		\label{fig: example simulation short}
		
	\end{figure*}

	\begin{figure}
		
		\centering
		\pic[0.8]{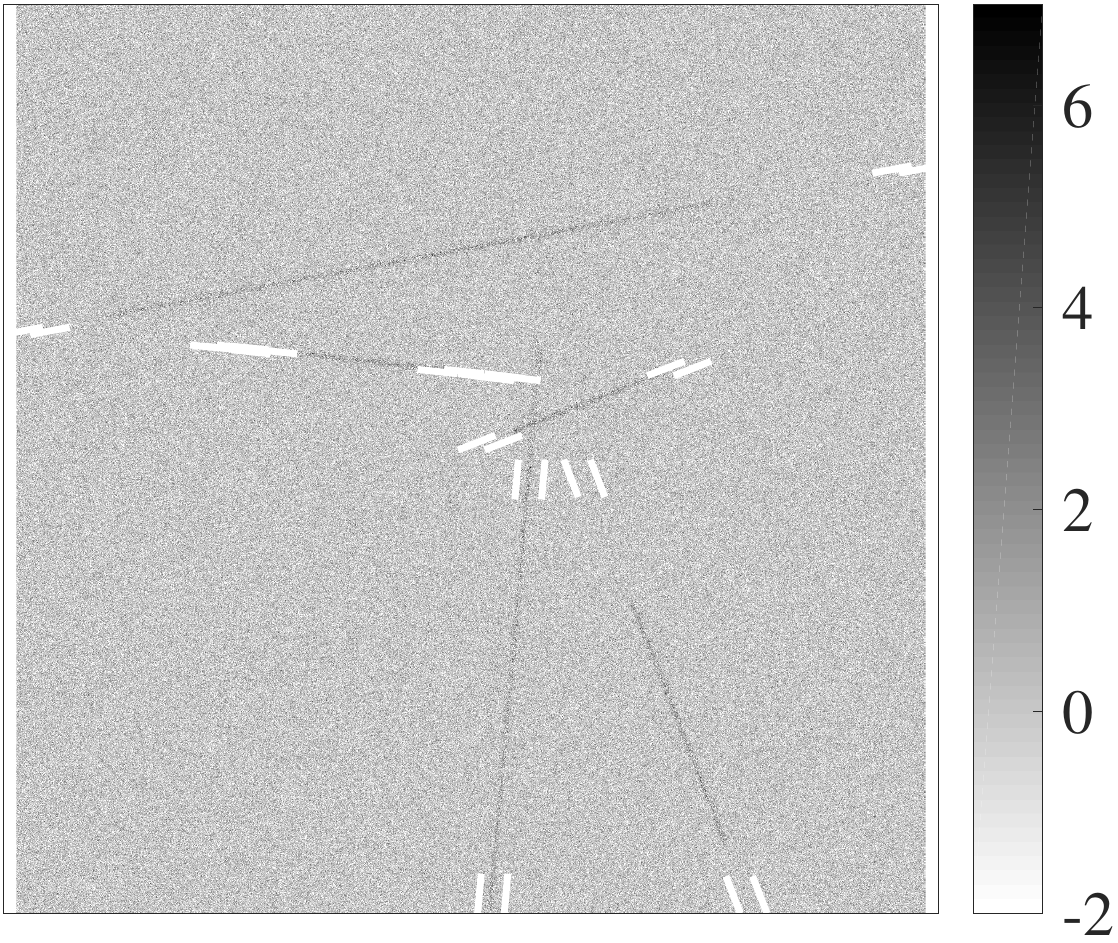}
		\caption{A set of 5 randomly generated streaks, 
			with white guidelines marking the beginning and end points 
			of all streaks found by	the multi-streak pipeline. 			     
		}
		\label{fig: example simulation multi}
		
	\end{figure}
	
	To check the efficiency of the FRT streak detection algorithm, 
	we simulated streaks in various angles, 
	start and end points, and levels of intensity. 
	In each row of the image, we find the pixel center 
	that is nearest to where the line passes through that row, 
	and set that pixel value to the chosen intensity.
	For lines with angles above $45^\circ$, we do the same but for each column. 
	This results in a digital approximation to a straight line of single pixel width. 
	We then convolve the image with a PSF 
	(in our simulations, we used a 2D Gaussian with $\sigma_p=2$ pixels).
	We add white Gaussian noise to all pixels (with variance equal to one). 
	For an intensity per unit length $\xi$,
	each pixel that the line crosses is set to $\xi / g(\theta)$, 
	where $g(\theta)=\max(|\cos\theta|,|\sin\theta|)$ is a geometric factor
	that is added because diagonal lines have longer sections in each pixel.  
	We calculate the expected $S/N$ from such a simulation using Equation~\ref{eq: total snr}.
	We run the streak detection code on the simulated images
	and measure the position and value of the 
	maximum of the normalized Radon image $\tilde S$ from Equation~\ref{eq: radon snr}. 
	The position of the peak gives the initial position $x_0$ of the streak
	and the slope parameter $\Delta x$, which is readily converted to the streak angle $\theta$. 
	The value of the maximum gives the detection $S/N$ directly. 
	
	An example of a very bright simulated streak and the resulting Radon image is shown in Figure~\ref{fig: example simulation bright}. 
	The algorithm succeeds in recovering the correct streak parameters. 
	The algorithm recovers on average more than 95\% of the expected\footnote{
		For an ensemble of simulated streaks with different noise realizations, 
		the recovered $S/N$ fluctuates around the mean value with a standard deviation of 1.}
	$S/N$. Presumably, this information loss is due to 
	the discretization process and edge effects. 
	
	We simulate a faint streak with a theoretical $S/N$ just above the threshold for detection. 
	The results are shown in Figure~\ref{fig: example simulation faint}. 
	Even though the streak is barely visible by the eye 
	(the brightest pixels of the streak contain a signal that is 0.12 times lower than the noise RMS)
	the streak is still detectable with a $S/N\approx 8.6$. 
	
	We test the detection of short streaks using the same pipeline, 
	activating the short-streak detection function in the FRT algorithm. 
	An example of a simulated short streak is shown in Figure~\ref{fig: example simulation short}. 
	The algorithm detects the streak and finds the length of the streak to within a factor of two. 
	Note that this streak, of length $L=72$ pixels, 
	is too faint to be reliably detected using the regular FRT pipeline.
	Without the short streak detection function, 
	this streak can sometimes be detected with a $S/N\approx 6$, 
	but in many cases it is overpowered by noise in the image and is not identified at all. 
	Using the short-streak pipeline, 
	the streak is detected with a $S/N = 13$. 
		
	To test the streak detection pipeline on multiple streaks,
	we generated five random streaks 
	with intensities per unit length in the range 3-10 times the image noise RMS 
	and used the iterative method described in \S\ref{subsec: multi}. 
	An example image where all the streaks have been detected is shown in Figure~\ref{fig: example simulation multi}. 
	We ran this simulation multiple times with random streak lengths and positions. 
	This method does not always detect all five streaks, 
	especially if some of them are very faint 
	and intersect with brighter streaks. 
	In most cases, the majority of streaks are correctly detected. 
	
	
	\section{Tests on real images}\label{sec: real data}
	
	\begin{figure*}
		\centering
		\pic[0.8]{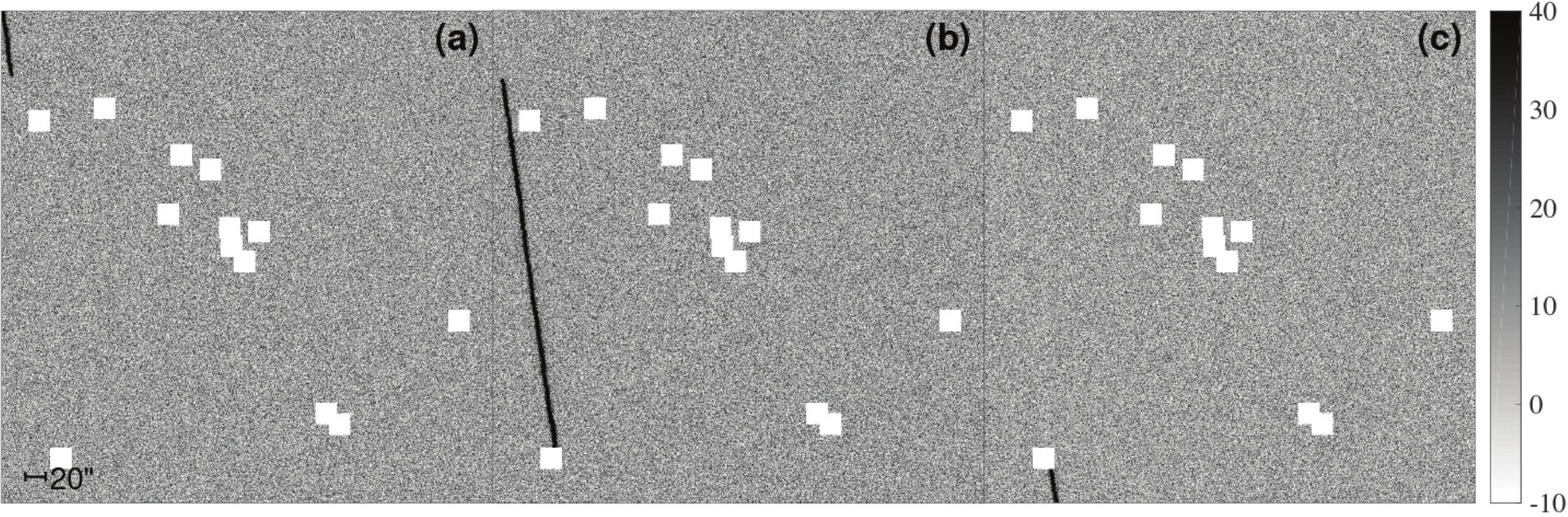}
		\caption{A bright streak from a LEO object, 
			recorded over three frames of 1 second exposures.
			Color scale is inverted (black is bright). 
			White squares are cutouts around stars. 
			The streak is very bright with 
			$\approx 300$ counts/pixel along the center of the line, 
			compared to a noise standard deviation of $\approx 22$ counts/pixel,
			and a $\sigma_p\approx 2$ pixels 
			(or $S/N\approx 55$  per resolution element) 
			and can be detected easily by any method.
			The algorithm shows good results in finding the angle and intercept of the lines. 
		}
		\label{fig: example strong streak}
	\end{figure*}
	
	\begin{figure}
		\centering
		\pic[0.8]{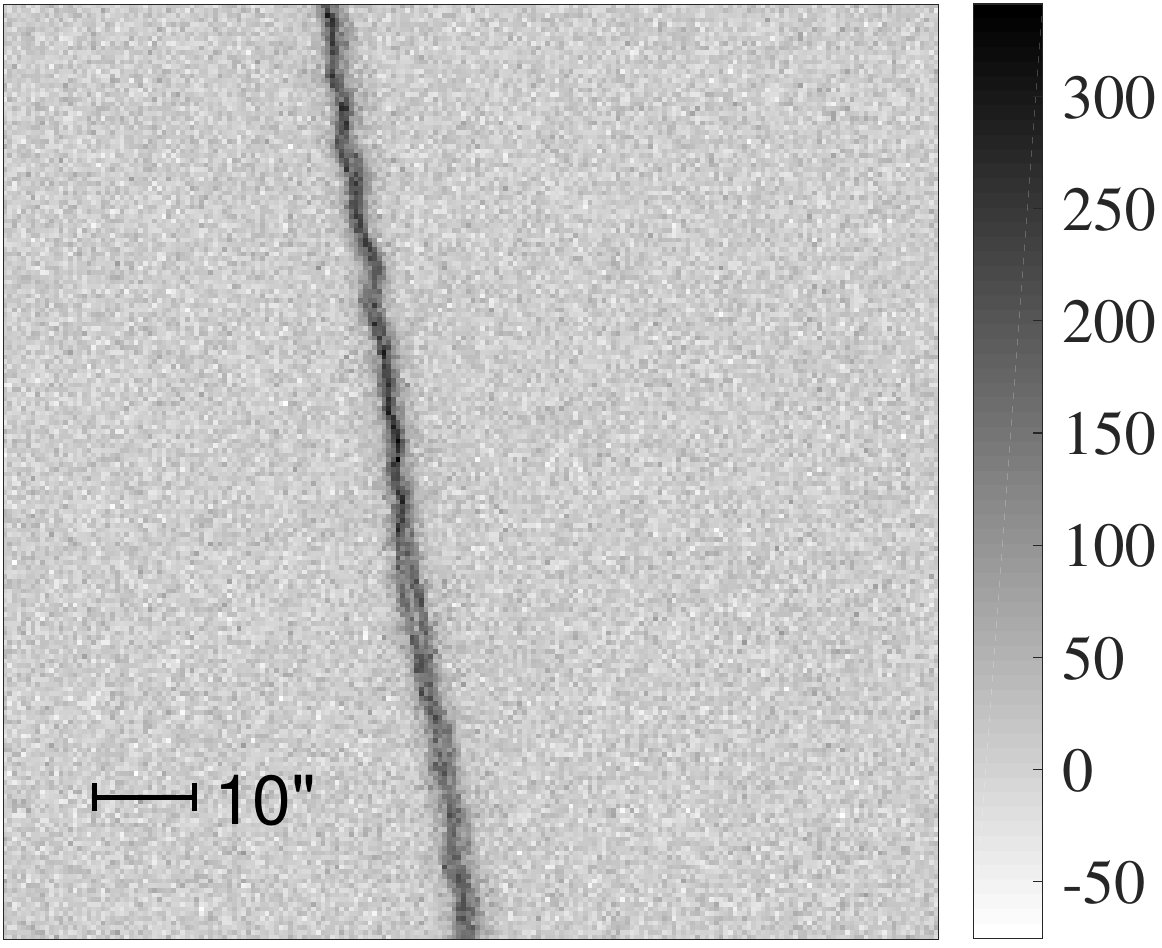}
		\caption{The same streak as shown in Figure~\ref{fig: example strong streak}b
			enlarged around the center of the streak. 
			The wobbles in the streak are likely caused by tracking errors 
			and more importantly by the turbulent atmosphere (astrometric scintillation).
			The wobble length scale is about 10 arcseconds. 
			Assuming the atmospheric scintillation time scale of about 10\,ms, 
			the expected angular speed of the object is on the order of 1000 arcsecond/second. 
			This is close to the actual angular speed of 788 arcsecond/second, 
			measured from the length of the streak (in Figure~\ref{fig: example strong streak}b)
			divided by the exposure time. 
		}
		\label{fig: example wobble}
	\end{figure}
	
	\begin{figure*}
		\centering
		\pic[0.8]{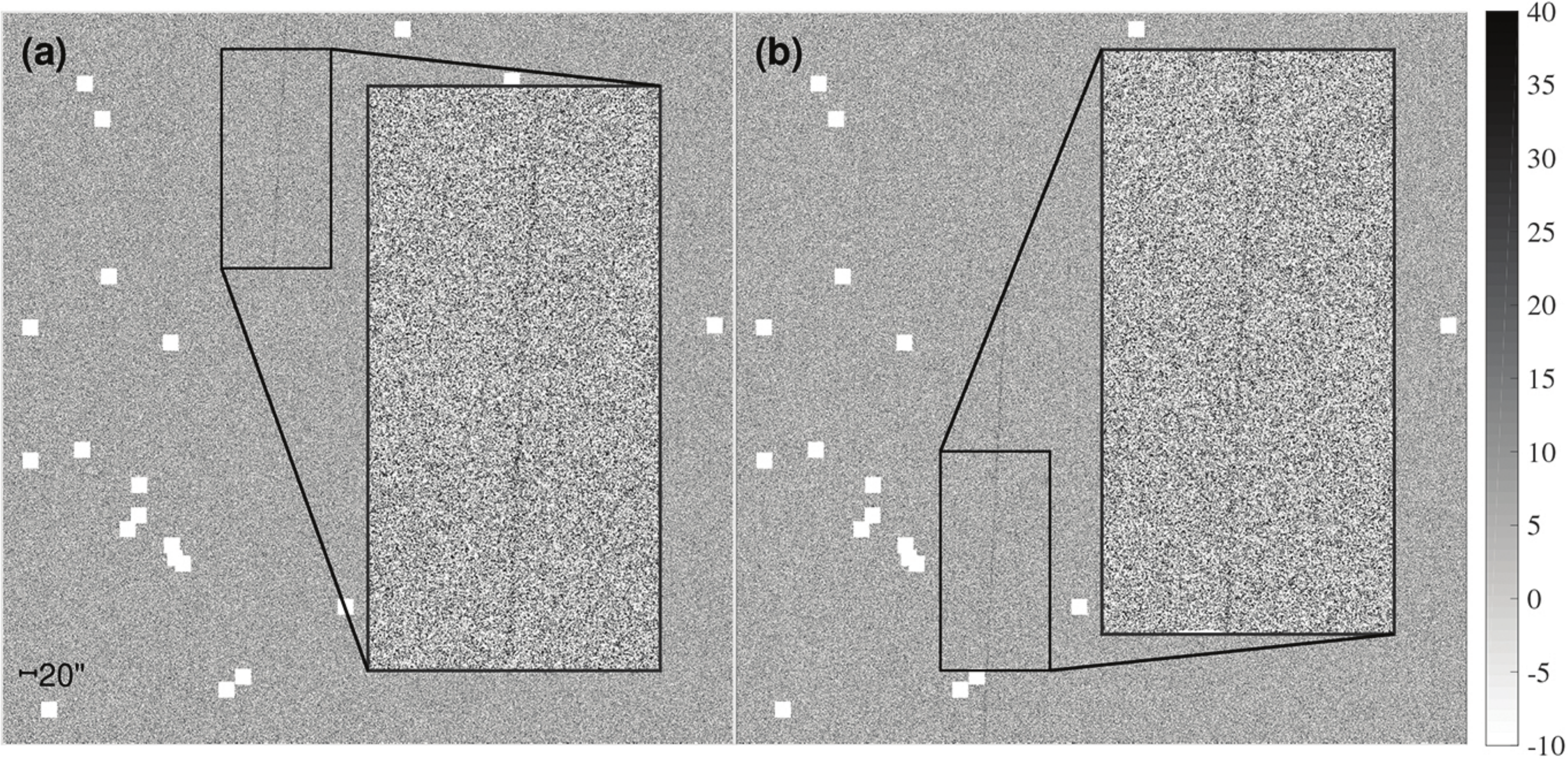}
		\caption{A faint streak from a LEO object, 
			recorded over two frames of 1 second exposures.
			Color scale is inverted (black is bright). 
			White squares are cutouts around stars. 
			This streak is barely visible to the eye, 
			with the peak pixel intensity lower than the noise standard deviation. 
			The calculated $S/N$ per resolution element is $\approx 1.3$. 
			(a) the first frame where the streak is detected, with an inset magnifying a part of the streak. 
			This streak was detected with an integrated $S/N$ of $13$. 
			(b) the second frame where the streak is detected, with an inset magnifying a part of the streak. 
			This streak was detected with an integrated $S/N$ of $19$. 
		}
		\label{fig: example faint streak}
	\end{figure*}
	
	\begin{figure}
		
		\centering
		
		\pic[0.8]{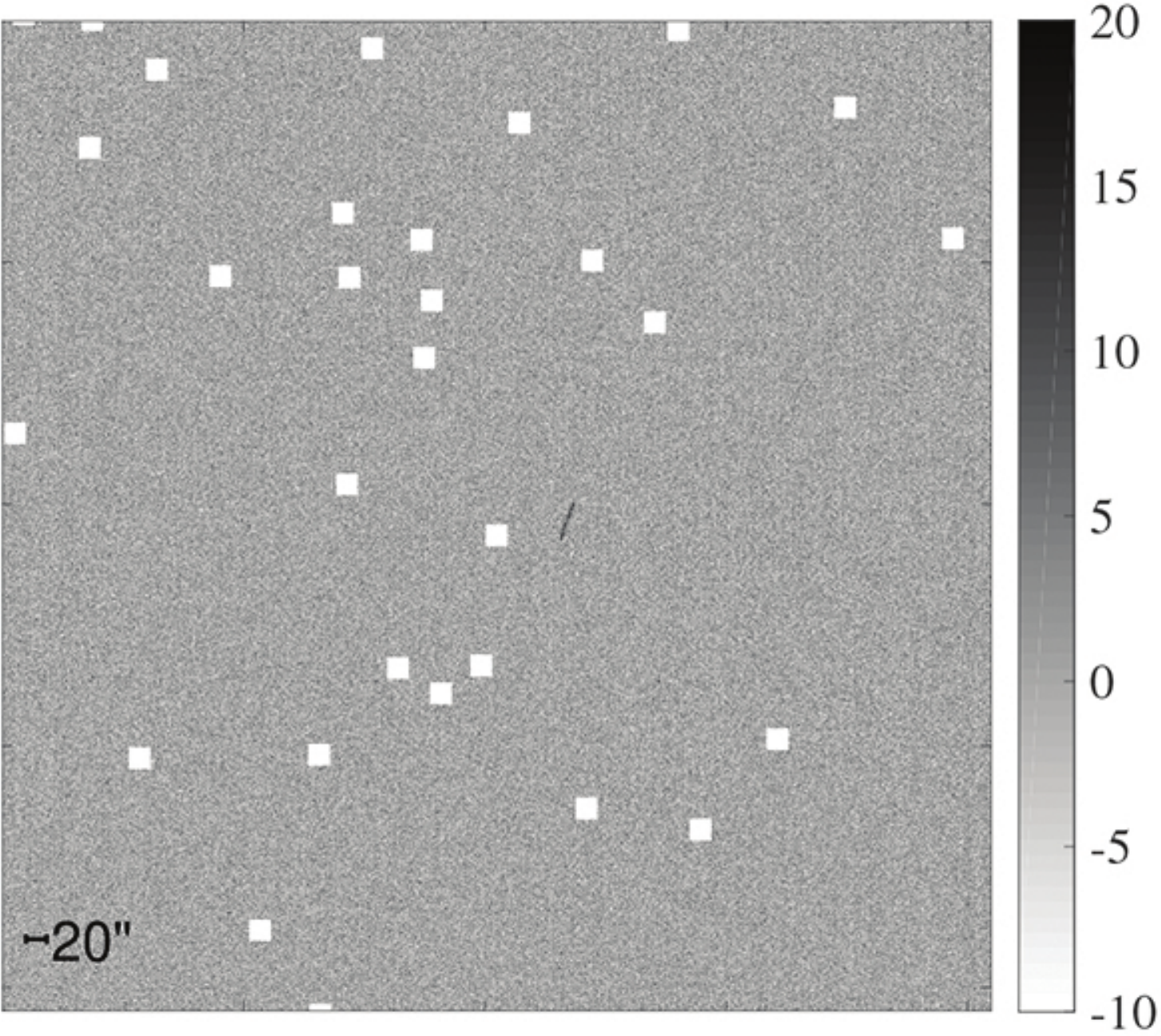}
		\caption{Example GPS streak detected using the short streak detection code presented in \S\ref{subsec: short}. 
			Color scale is inverted (black is bright). 
			White squares are cutouts around stars.  
			The image noise standard deviation is estimated at $\approx 8.5$ counts/pixel, 
			while the streak brightness peaks at $\approx 10$ counts/pixel, 
			and the PSF width is $\sigma_p\approx 2$ pixels,
			equivalent to a $S/N\approx 5$ per resolution element. 
			This streak is detected with a total $S/N=18.6$. 
		}
		\label{fig: example GPS}
		
	\end{figure}
	
	\begin{figure}
		
		\centering
		
		\pic[0.7]{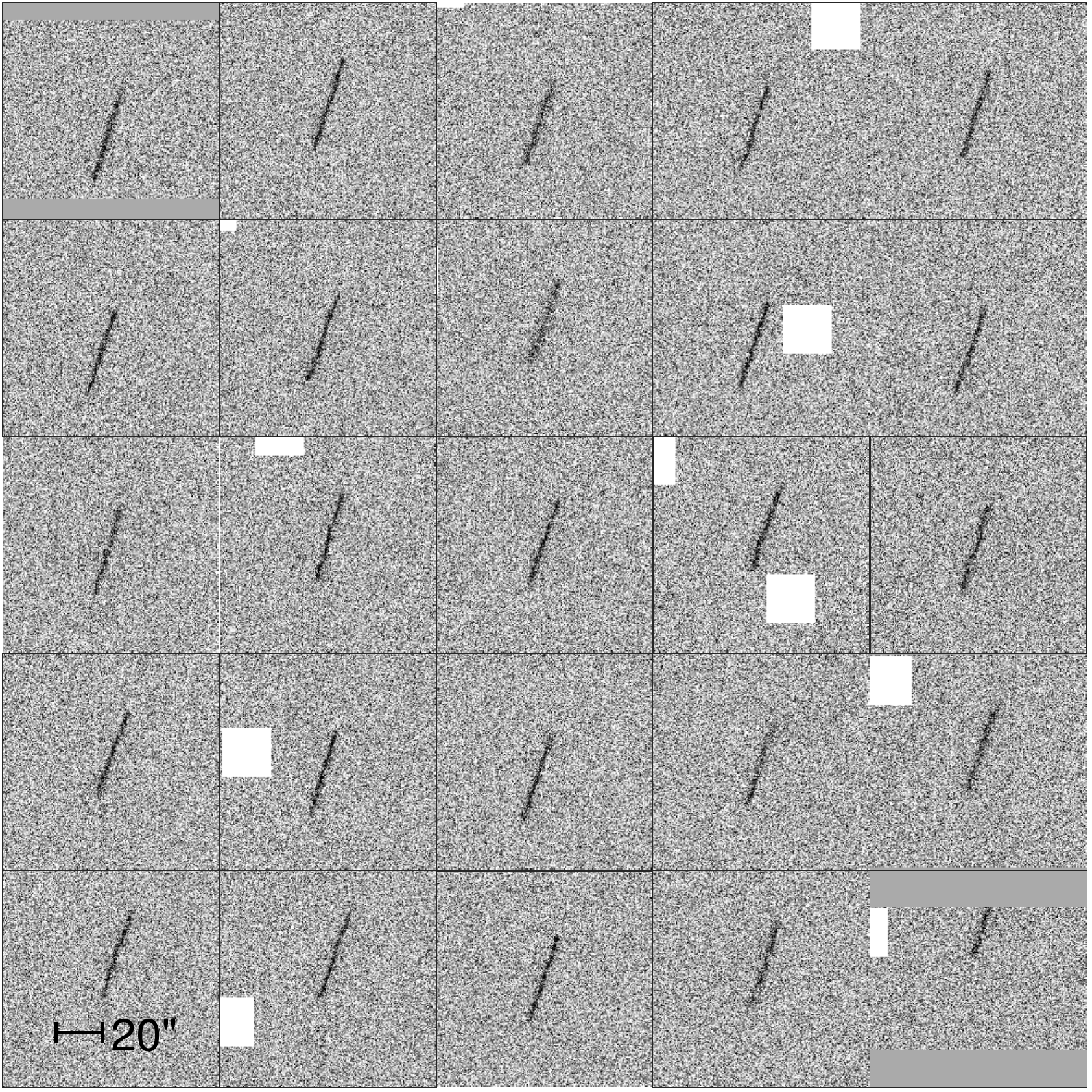}
		\caption{Example GPS streaks detected using the short streak detection code presented in \S\ref{subsec: short}. 
			Close cropping around each streak in each separate frame is shown for 25 out of 26 images where this satellite is detected. 
			Color scale is inverted (black is bright). 
			White squares are cutouts around stars. 
			Some of the images show streaks that are partially removed when crossing near stars, 
			but are still detected. 
		}
		\label{fig: example GPS cutouts}
		
	\end{figure}
	
	\begin{figure*}
		\centering
		\pic[0.8]{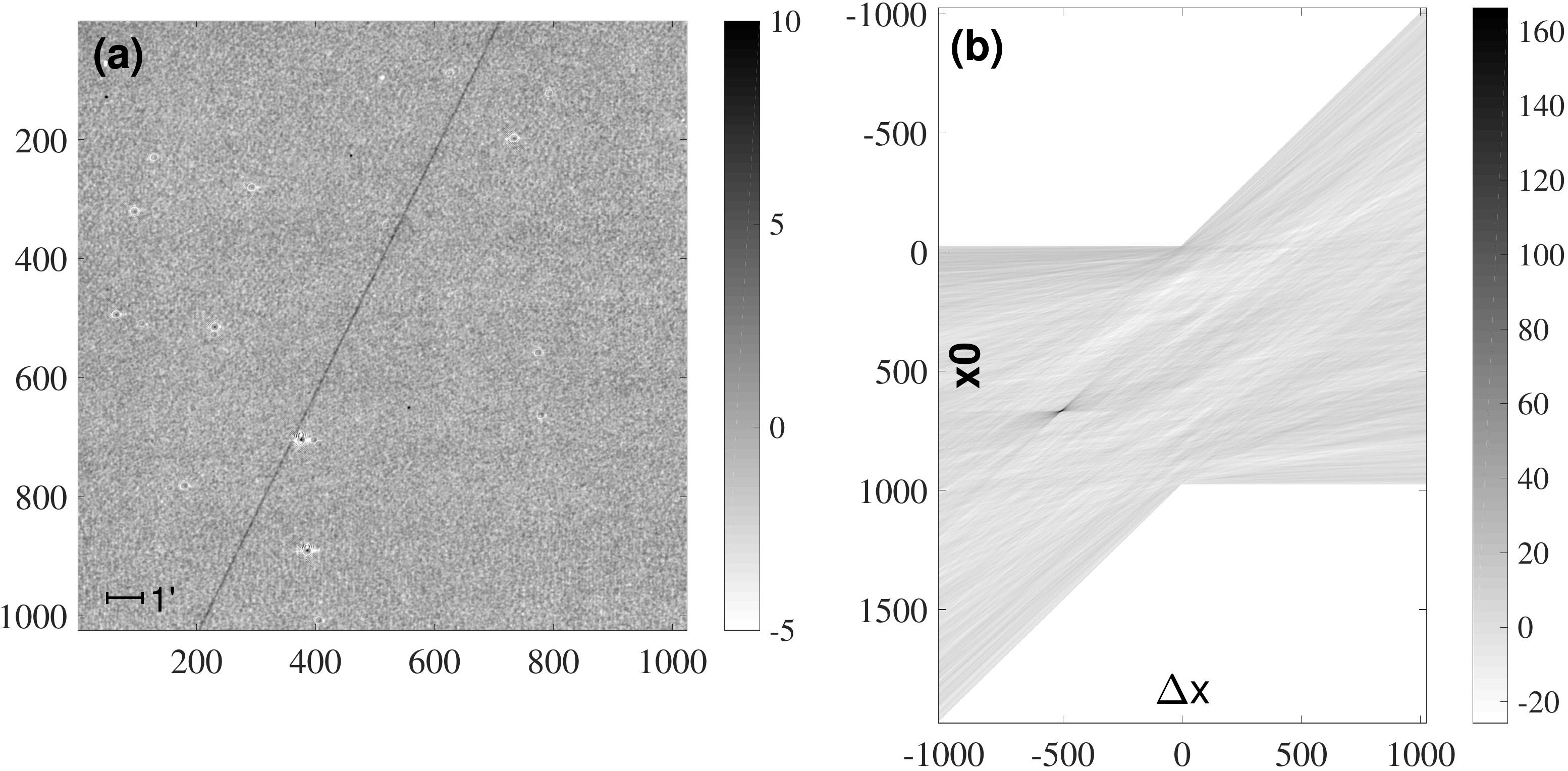}
		\caption{A strong streak from the Palomar Transient Factory. 
			The image has undergone pre-processing (image subtraction) 
			and the streak with a
			$S/N\approx 11$ per resolution element
			is easily detected at a $S/N=175$ using the FRT algorithm. 
			Runtime for streak detection on this $1024\times 1024$ pixel image
			is~$\approx 0.9$ seconds. 
		}
		\label{fig: example ptf bright}
	\end{figure*}
	
	We tested the FRT streak detection pipeline
	on real images from the Kraar observatory (\S\ref{subsec: kraar}) 
	and from the Palomar Transient Factory (\S\ref{subsec: ptf}).
	
	\subsection{Observations at the Kraar observatory}\label{subsec: kraar}
	
	We performed observations at the Kraar observatory at the Weizmann Institute
	for approximately two hours each night for two nights,
	on 2017 July 20 and August 8.
	Images were taken 
	using a Meade 40\,cm f/10 telescope, an Andor Zyla 5.5 sCMOS camera,
	a focal reducer (converting to f/7) and a $V$ filter. 
	The pixel scale was $0.47''$/pixel and 
	we read a $2048\times2048$ pixel subsection of the sensor\footnote{
		The Zyla detector has $2160\times2560$ pixels, but we chose to crop the frame to an integer power of 2.}.
	To mitigate sensor line-noise, we exclude the range $-50<x<50$ from the Radon map, 
	effectively ignoring lines up to an angle of $|\theta|<1.4^\circ$ from the $y$ axis.\nolinebreak
	
	We observed fields in which GPS satellites were predicted to pass\footnote{{Using an online tool: https://projectpluto.com/gps\_find.htm}}.
	Images were taken during the first two hours after sunset. 
	For each target, we positioned the telescope a few minutes ahead of time
	and took 400 or 500 images at 1 second exposure time. 
	
	Low Earth Orbit (LEO) satellites and space debris are expected to move
	at more than 1000 arcsecond/second, 
	but GPS satellites have a slower angular velocity of 30-40 arcsecond/second. 
	Exposure time was chosen ($T=1$\,s) to make sure the GPS satellites are still streaked
	and so that LEO objects would leave a large part of their streaks inside the frame. 
	At such exposure times, the images are background dominated 
	and no further increase in the $S/N$ is gained by taking longer exposures. 
    
	We ran a blind search over all images taken. 
	We performed a visual inspection of the images on which the algorithm was triggered, 
	and adjusted the pipeline parameters 
	to remove cosmic rays and stars that still remained outside the cut-out region. 
	Some examples of the detected streaks are presented. 
	A strong streak that crosses the corner of the field of view in three frames is presented in Figure~\ref{fig: example strong streak}.
	This streak's intensity is high enough 
	to be easily detected by any transient detection method
	($\approx 300$ counts/pixel along the center of the line, 
	with a noise standard deviation of $\approx 22$ counts/pixel, 
	and a PSF width of $\approx 2$ pixels, 
	which translates to a $S/N\approx 55$ per resolution element). 
	We note that the pipeline extracted the correct properties of the line 
	(slope, intercept and length).
	
	We can estimate, based on the streak brightness, 
	that the $S/N$ for detection should be $\approx 1000$.
	However, the algorithm only recovers a $S/N=306$. 
	This is due to two reasons:
	(a) the streak is not aligned perfectly with the partitions of the data to powers of two;
	(b) the streak is not exactly a straight line, but wobbles as it crosses the frame, as seen in Figure~\ref{fig: example wobble}.	
	The first problem is unavoidable when using our version of short-streak detection, 
	but can be addressed after detection by brute force integration. 
	The second problem is due to astrometric scintillation noise and tracking errors.
	Once more, the $S/N$ can be estimated more accurately after detection 
	by running a filter that tracks the wobbles of the line (e.g., Kalman filter; \citealt{Kalman_filter_1960}). 	
	If the streak begins or ends outside the frame, 
	its velocity cannot be inferred from its length. 
	In some cases, however, the length scale of the wobbles 
	can be used to estimate the angular velocity of the streak. 
	For example, we expect that the wobble time scale will be 
	of the order of the scintillation time scale 
	(e.g., the telescope diameter divided by the wind speed).
	 
	A more challenging example is shown in Figure~\ref{fig: example faint streak}, 
	where the algorithm is able to detect, in two consecutive frames, 
	a faint streak that is barely visible to inspection by the eye. 
	This streak has a $S/N$ per resolution element of approximately $1.3$, 
	but the integrated signal-to-noise of these detections (independently for each consecutive frame) 
	is $S/N\approx 13$ and $19$, due to the difference in length of the streaks.
	The detection threshold was set at $S/N=10$. 
	It should, therefore, be possible to detect even fainter streaks, as was done in the simulations. 

	Although we find several LEO streaks from satellites or space debris, 
	our primary target was GPS satellites, 
	which create shorter streaks in the image
	and can be detected only when applying the short-streak-detection code presented in \S\ref{subsec: short}.
	We targeted five GPS satellites during our first test night, 
	and succeeded in detecting two targets. 
	Each satellite was observed in several consecutive frames. 
	We present an example image showing the relative size and brightness of one GPS streak in Figure~\ref{fig: example GPS}
	and more detailed cutouts around the found streak positions of the remaining 25 out of 26 frames in Figure~\ref{fig: example GPS cutouts}.
	

	\subsection{Identifying streaks in archival PTF data}\label{subsec: ptf}
	
	We tested the FRT pipeline on an image from 
	the Palomar Transient Factory (PTF, \citealt{palomar_transient_factory_Law_2009}).
	The images were reduced as described in \citep{photometric_calibration_PTF_Ofek_2012, image_processing_PTF_IPAC_Laher_2014} 
	and reference subtraction was performed using the image subtraction algorithm\footnote{We used code from the MATLAB astronomy and astrophysics toolbox in \citealt{matlab_package_Ofek_2014}.}
	 of \cite{Subtraction_Zackay_2016}.
	In this image, a strong streak is visible. 
	Even though the image has undergone image subtraction, 
	some artefacts remain, due to saturated stars and cosmic rays. 
	Still, the streak is easily detected with a $S/N=175$ and is shown in Figure~\ref{fig: example ptf bright}. 
	
	
	\section{False alarm rate}\label{sec: false alarm}
	
	\begin{figure}
		
		\centering
		
		\pic[1]{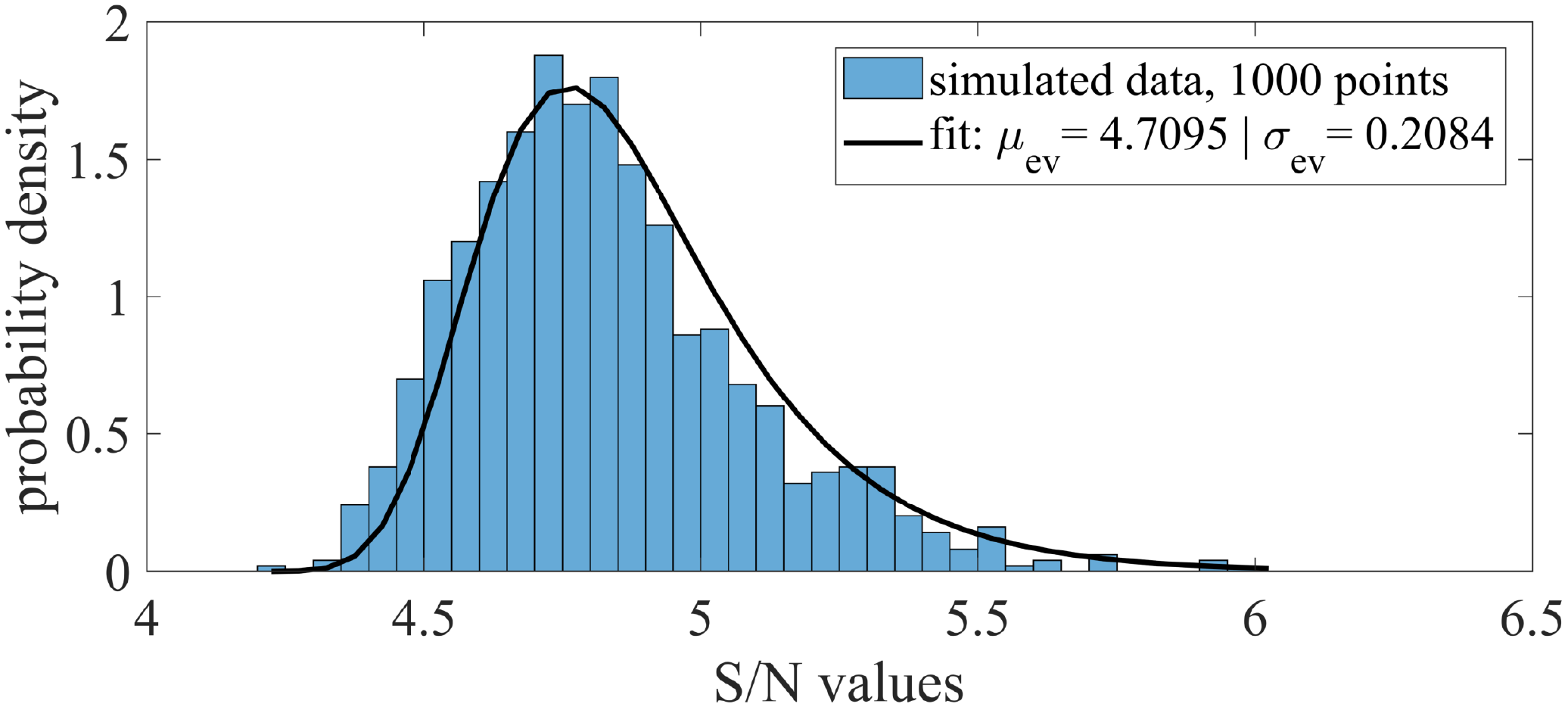}
		
		\caption{Signal-to-noise ratio results for background-only simulated images. 
			Each pixel in the Radon space, after normalizing by the Radon variance image,
			is normally distributed, with $\mu=0,\sigma=1$. 
			The maximum of these pixels is described by an extreme-value distribution,
			with the parameters of the fit plotted over the simulated values. 
			Such a fit can be used to find a threshold for each set of observations. 
		}
		\label{fig: background snr sim}
		
	\end{figure}
	
	\begin{figure}
		
		\centering
		
		\pic[1]{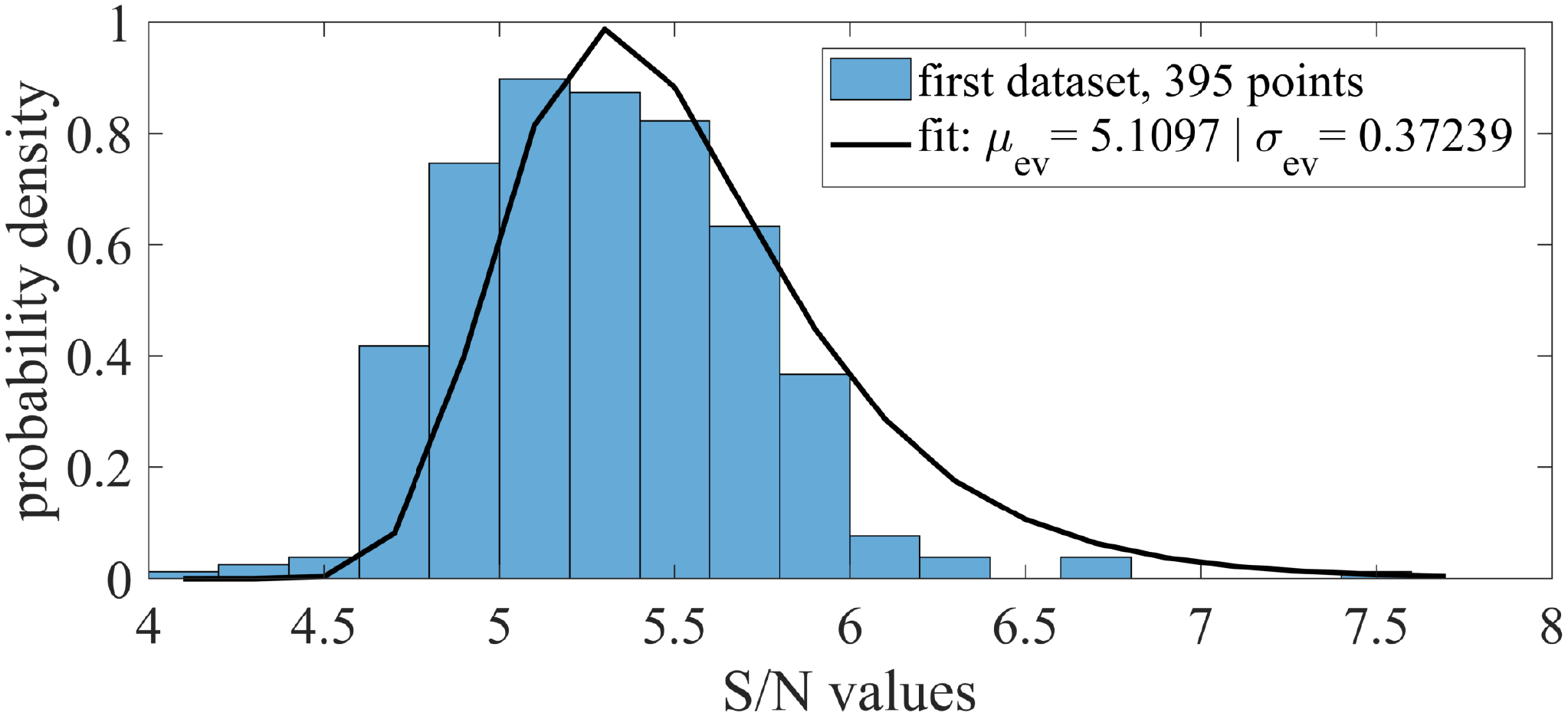}
		
		\caption{Signal-to-noise ratio results for background-only images from the first set of observations 
			from the Kraar observatory at the Weizmann Institute. 
			The resulting $S/N$ values from 400 images are plotted, removing 5 images that contained streaks. 
			The extreme-value fit to the results can be used to find a threshold for this set of observations. 
		}
		\label{fig: background snr data}
		
	\end{figure}
	
	It is useful to determine a threshold $S/N$ for detection
	that is as low as possible, given some acceptable false alarm probability. 
%
%
%
%
	We determine the threshold 
	by measuring the detected maximal $S/N$ of all the images that do not have detectable streaks. 
	The peak $S/N$ of noise-only images is the maximum of the normally distributed pixels in Radon space.
	Therefore, the values of the maxima of a batch of Radon images should follow an extreme value distribution 
	(i.e.,~the Gumbel distribution, see~\citealt{extreme_value_distributions_Kotz_2000}).
	For noise-only images of size $2048\times 2048$, 
	we expect to find most images have a maximal $S/N\sim 5$. 
	After performing a fit to the extreme-value distribution, 
	we use the CDF of the distribution to determine the appropriate threshold 
	for the required false alarm rate. 
	If the distribution of noise-only images does not follow this distribution 
	or has a mean value that is too high or too low, 
	it may be an indication of poor pre-processing of the images or some other problem in the pipeline. 
	
	The $S/N$ results for simulated noise-only images
	are shown in Figure~\ref{fig: background snr sim}.
	The results follow an extreme-value distribution, 	
	\begin{equation}\label{eq: evpdf}
		f(x; \mu_\text{ev}, \sigma_\text{ev})=\frac{1}{\sigma_\text{ev}} \exp\left[ -\frac{x-\mu_\text{ev}}{\sigma_\text{ev}} - \exp\left( -\frac{x-\mu_\text{ev}}{\sigma_\text{ev}}\right)\right],
	\end{equation}
	where the $S/N$ is represented by $x$, the position parameter $\mu_\text{ev}= 4.7$ and the scale parameter $\sigma_\text{ev}=0.216$.
	These results may be used, along with the extreme-value cumulative distribution function,
	to find a threshold with a given false alarm rate. 
	When using the short-streak method, 
	the fit parameters are only slightly different, 
	with $\mu_\text{ev}=5.03$ and $\sigma_\text{ev}=0.206$. 
	
	The noise-only images from the first set of observations (of 400 images) 
	can be used to determine the noise properties of the data and to find the desired threshold. 
	We removed all the images with $S/N>10$, corresponding to 5 images where a streak was detected.
	We fit the remaining $S/N$ values using 
	an extreme-value distribution. 
	The results are shown in Figure~\ref{fig: background snr data}. 
	Using the inverse CDF of the extreme-value distribution given in Equation~\ref{eq: evpdf}, 
	we find that a threshold of $\cong 10.25$ 
	will have a single false positive result for every $10^6$ such images. 
	
	
	\section{Code}\label{sec: code}
	
	The streak detection code is available online in Python and MATLAB versions. 
	
	The code for streak detection includes three classes and one function, 
	along with some utility functions. 
	The function performs the FRT on images, 
	with optional arguments for zero padding, expanding the sides of the matrix, and transposing the matrix. 
	This allows users to directly apply the FRT to images without actively searching for streaks. 
	
	To search for streaks, the \emph{Finder} class is given as an optional argument to the FRT function. 
	The Finder calculates the Radon variance map 
	and can perform a cross-correlation with a given PSF before using the FRT. 
	Inside the FRT function, it checks for streaks that pass the threshold in the full Radon image or in sub-frames, 
	and saves the results as \emph{Streak} class objects. 
	The Finder can find multiple streaks or short streaks, depending on the Finder object parameters. 
	The Streak class contains all measured data about the detected line, 
	and translates the raw Radon coordinates to physical coordinates in the original image. 
	A third class is the \emph{Simulator}, which produces streaks for testing the Finder capabilities.
	
	This software is provided, along with some auxiliary functions and documentation, in Python\footnote{\url{https://github.com/guynir42/pyradon.git} } 
	and MATLAB\footnote{\url{https://github.com/guynir42/radon.git}}.
	It is also available as part of the MATLAB Astronomy \& Astrophysics toolbox\footnote{\url{https://webhome.weizmann.ac.il/home/eofek/matlab/}
		\url{}} \citep{matlab_package_Ofek_2014}.
	
		
	\pagebreak
		
	\section{Summary}\label{sec: summary}
	
	Linear features in astronomical images can be generated by image artefacts (e.g.,~diffraction spikes);
	some cosmic ray hits; 
	or fast moving objects, such as asteroids, satellites and space debris. 
	The requirement for an efficient, fast, and reliable streak detection method 
	becomes more pronounced as surveys cover larger fields of view with higher sensitivity. 
		
	The FRT algorithm is a fast and optimal way to detect linear features in astronomical images. 
	Detecting streaks can be used either to remove them or flag the affected pixels, 
	or to detect those fast moving objects themselves. 

	We demonstrate the efficiency of using a PSF matched-filter followed by a Radon transform
	as a way to find faint streaks. 
	We present an efficient implementation of the Radon transform 
	and extend it to finding short streaks and multiple streaks. 
	We show that the distribution of results for noise-only images 
	follows a well understood extreme-value distribution 
	and show how a detection threshold with a predefined false-positive rate 
	can be chosen directly from the data. 
	We test our streak-detection pipeline on simulated streaks and real data 
	and show that the algorithm succeeds in detecting and identifying streak parameters 
	of faint streaks. 
		
	\section*{Acknowledgments}
	
	G.N.~would like to thank H.~Horn for her participation in observations 
	and O.~Springer for his insights regarding signal and noise. 
	B.Z.~acknowledges the support from the infosys fund.
	E.O.O.~is grateful for the support by
	grants from the 
	Israel Science Foundation, Minerva, Israeli Ministry of Science,
	the US-Israel Binational Science Foundation,
	and the I-CORE Program of the Planning
	and Budgeting Committee and the Israel Science Foundation.

	\bibliographystyle{aasjournal}	
	\bibliography{references}
	
	\pagebreak
	
	\appendix
	\section{Optimal statistic for streak detection}~\label{sec: likelihood ratio}
	
	\newcommand{\cev}[1]{\reflectbox{\ensuremath{\vec{\reflectbox{\ensuremath{#1}}}}}}
	
	A streak $s$ in an astronomical image can be modeled 
	by a straight line convolved with the PSF of the image. 
	The image $M$ can be modeled by a streak with added Gaussian, i.i.d noise:
	\begin{equation}\label{eq: streak and image model}
		s = \xi\ell(x_1,y_1,x_2,y_2) \otimes P \qquad,\qquad M = s + \varepsilon.
	\end{equation}
	Here $\xi$ is the intensity per unit length of the streak, 
	$\ell$ is the single-pixel width line going from coordinates $(x_1,y_1)$ to $(x_2,y_2)$, 
	$P$ is the the PSF of the system,
	$\otimes$ is the 2-dimensional convolution operator
	and $\varepsilon$ is the background noise, which is assumed to be Gaussian, independent and identically distributed (i.i.d). 
	We also assume the images are background-noise dominated.
	The optimal way to detect such objects can be derived
	using the Nyman-Pearson lemma~\citep{max_likelihood_Neyman_Pearson_1933}, 
	which states that the most powerful test\footnote{Most powerful test in the sense that for any given detection threshold the false alarm probability is minimal.} 
	for differentiating two simple hypotheses is using the likelihood ratio test:
	\begin{equation}
		\Lambda \equiv \frac{\mathcal{L}(d|\mathcal{H}_0)}{\mathcal{L}(d|\mathcal{H}_1)}<\eta,
	\end{equation}
	where the ratio of the likelihood, $\mathcal{L}$, of the data $d$ given the null hypothesis, $\mathcal{H}_0$, 
	to the likelihood of the data given the alternative hypothesis, $\mathcal{H}_1$, is compared to
	some threshold $\eta$. 
	In this case, the null hypothesis is that of an image with no streak, 
	but only with Gaussian, uncorrelated and independent, equal variance noise:
	\begin{equation}
		\mathcal{H}_0: M_q = \varepsilon(0, \sigma).
	\end{equation}
	Here, $M_q$ are the background subtracted image data points with the grid coordinate $q$, 
	and $\varepsilon$ is Gaussian noise with zero mean and variance $\sigma^2$. 
	The alternative hypothesis is of a streak imposed on top of the noise:
	\begin{equation}
		\mathcal{H}_1: M_q = s_q(\Theta) + \varepsilon(0, \sigma),
	\end{equation}
	where $s_q$ is the intensity of the streak in each pixel (as in Equation~\ref{eq: streak and image model}), 
	and the length, angle, position, intensity and PSF of the streak is represented by $\Theta$. 
	In this case, the likelihood of measuring the data points $M_q$ given each hypothesis is:
	\begin{align}
		\mathcal{L}(M_q | \mathcal{H}_0) &= \frac{1}{\sqrt{2\pi}\sigma}\exp\left[-\frac{\sum_q M_q^2}{2\sigma^2}\right] ,\quad \\
		\mathcal{L}(M_q | \mathcal{H}_1) &= \frac{1}{\sqrt{2\pi}\sigma}\exp\left[-\frac{\sum_q (M_q-s_q)^2}{2\sigma^2}\right],
	\end{align}
	and taking the log of the likelihood ratio: 
	\begin{align}
		\log\Lambda(\theta) &= \frac{1}{2\sigma^2}\sum_qM_q^2 - \sum_q(M_q-s_q(\theta))^2 \notag\\
							&= \frac{1}{2\sigma^2}\sum_q 2M_q s_q(\theta) - s_q(\theta)^2.
	\end{align}
	This expression is further simplified by absorbing into the threshold the constant terms $2\sigma^2$ and $s_q^2$, 
	which do not depend on the data. 
	Therefore, the most powerful test for the detection of any known distribution of intensity (in this case, streaks) 
	is given by
	\begin{equation}
		\log \Lambda(\theta) = M_q s_q(\theta) = M\otimes \cev{s}, 
	\end{equation}
	where $\cev{s}$ is $s$ with flipped coordinates. 
	Thus, the statistic is the cross-correlation of the data $M$ with the filter of shape $s$.
	This test is optimal for any choice of $s_q$, including streaks of varying intensity,
	or for arbitrarily shaped sources, as long as the shape of the source is known. 
	In this work, we assume the light is distributed along a straight line that is widened by a 2-dimensional Gaussian PSF, 
	and that the streak intensity is constant along the line.
	
	This test is the most powerful test if the noise is normally distributed.
	For other cases, different tests should be used 
	(e.g.,~an optimal test for the case of Poisson-distributed noise was derived in~\cite{match_filter_Poisson_noise_Ofek_Zackay_2017}).
	
	\vspace{1cm}
	
\end{document}